\documentclass[10pt,twocolumn,aps,pra,floatfix]{revtex4-1}
\usepackage{graphicx} 
\usepackage{bm}
\usepackage{hyperref} 
\usepackage{siunitx}
\usepackage{amsmath}
\usepackage{upgreek}
\usepackage{multirow}
\usepackage[T1]{fontenc}

\begin{document}

\title{Determination of Quantum Defects and Core Polarizability of Atomic Cesium via Terahertz and Radio-Frequency Spectroscopy in Thermal Vapor}
\author{Gianluca Allinson, Lucy A. Downes, C. Stuart Adams and Kevin J. Weatherill}
\date{\today}

\affiliation{$^1$Department of Physics, Durham University, South Road, Durham. DH1 3LE, United Kingdom.}
\date{\today}

\begin{abstract}
We present new measurements of quantum defects and core polarizabilities in cesium ($^{133}$Cs), based on transition frequency measurements between Rydberg states ($14 \leq n \leq 38$) obtained through terahertz (THz) and radio-frequency spectroscopy in a thermal atomic vapor. 
We perform a global fitting of our measurements to extract quantum defects of the $s_{1/2}$, $p_{1/2}$, $p_{3/2}$, $d_{3/2}$, $d_{5/2}$, $f_{5/2}$, $f_{7/2}$, $g_{7/2}$ and $g_{9/2}$ electronic states. Transitions between high angular momentum states ($4 \leq \ell \leq 8$) were measured to extract the Cs$^{+}$ dipole and quadrupole polarizabilities. We find $\alpha_{d} = 15.729(18)$~$a_{0}^{3}$ and $\alpha_{q} = 76.3(1.9)$~$a_{0}^{5}$ respectively. 
Using these results, and accounting for the covariances between parameters in the global fit, the energies for $n\ell_{j}$ Rydberg states can be estimated to a precision of a few MHz or less.
\end{abstract}

\maketitle

\section{Introduction}

High-resolution spectroscopy of atomic and molecular states is fundamental to our understanding of the structure of matter and allows us to refine fundamental theories and discover new physics~\cite{newphysics,Kozlov2018}. Detailed knowledge of the energy levels of atoms and molecules is also essential for the development of many modern atom-based quantum technologies~\cite{qmsensing,Saffman2010} with systems involving highly-excited Rydberg levels showing particular promise for quantum computing and electric-field-sensing applications~\cite{Adams_2020}. Rydberg atom-based systems are increasingly being used for electric-field sensing~\cite{Jing2020}, communications~\cite{Nowosielski:24,Nikunjkumap:22} and metrology~\cite{Sedlacek:12} applications spanning the radiofrequency~\cite{Lei:24} to terahertz range~\cite{Chen:22,Downes:20} and often include transitions to states with high angular momentum~\cite{highlE,Meyer:23}.
To model these systems, open-source software such as ARC~\cite{arc1} are widely used for calculating the properties of alkali atoms, relying on precision measurements of constants such as quantum defects to calculate energy levels, in order to improve predictive accuracy.\par
The most comprehensive measurements of the quantum defects in Cs were made in 1987 by Weber \& Sansonetti~\cite{Weber:87}. They determined the quantum defects of the $ns_{1/2}$, $np_{1/2}$ and $nd_{5/2}$ states, and rely on fine-structure intervals measured by Goy \textit{et\ al.}~\cite{Goy:82} for calculation of the $nd_{3/2}$ quantum defects, and from Sansonetti \& Lorenzen~\cite{Sansonetti:84} for calculation of the $np_{3/2}$ quantum defects. Data were also taken for the $nf_{5/2}$ and $ng_{7/2}$ states, with the latter having no other recent measurements. Recently, there have been more precise measurements for the $nf_{5/2, 7/2}$~\cite{Fqds} and the $ns_{1/2}$, $np_{1/2, 3/2}$ and $nd_{5/2}$ quantum defects~\cite{DeiglmayrCs} by millimetre-wave spectroscopy of ultra-cold Cs~\cite{DeiglmayrCs2}. However, these measurements still rely on data from older sources \cite{Weber:87} rather than providing an independent reference.
The most recent direct measurements for the quantum defects of the $nd_{3/2}$ state in Cs was made by Lorenzen \& Niemax in 1984 \cite{Lorenzen:84}. 
Consequently, there is no single independent reference for accurate values of the quantum defects for all states with $\ell\leq 4$. \par
The quantum defects of higher $\ell $ states ($\ell \geq 5$) can be calculated from the polarizability of the ionic core, as the wavefunction's penetration into the core is minimal ~\cite{Freeman}. The measurement of Rydberg high $\ell$ states has been used to extract ion polarizabilities~\cite{LUNDEEN2005161}. This technique is not limited to alkali metals~\cite{lithium,sodium,potas} and has been explored in alkaline-earth elements~\cite{Komara_2005,barium} and other species~\cite{nitrogen,sulfur}. For simpler species, such as He and H$_{2}$, comparisons between theoretical and experimental results have further validated these methods~\cite{PhysRevA.26.1228,PhysRevA.56.R4361}. Beyond this, ionic polarisabilities have a range of applications~\cite{Mitroy_2010} from the study of ion-neutral and atom-atom interactions~\cite{core_int_1,core_int_2,atom_atom} to addressing black-body radiation uncertainties in ionic and atomic clocks~\cite{StarkPol,bbratom}. Proposals to trap atoms using circular states~\cite{qc_circ} and improve measurements of the Rydberg constant have highlighted the need for improved ionic polarizability constants~\cite{circ1,circ2} with recent studies identifying it as a leading uncertainty~\cite{circ3}. Accurate determination of atomic polarizabilities is also critical for calculating tune-out wavelengths and Stark shifts, which depend on the contribution from the core~\cite{IonicTh1,TuneOut}. In a recent Cs measurement, this is the dominant source of error in determining the ratio of the 6$p$ reduced matrix elements~\cite{TuneOutCs}.
 \par
In this article, we present values for the quantum defects for the $s_{1/2}$, $p_{j}$, $d_{j}$, $f_{j}$ and $g_{j}$ states of Cs, using Rydberg electromagnetically induced transparency (EIT) in conjunction with terahertz (THz) and radio-frequency (rf) fields. 
Frequency intervals between Rydberg states are measured using coherent THz and microwave sources to couple the states. These results are compared to the most recent measurements of the Cs quantum defects. 
Finally, by using our measurements of Rydberg states with $\ell \geq 4$, accurate values of the Cs$^{+}$ dipole and quadrupole polarizabilities can be extracted. This enables precise calculations of states with $\ell \geq 5$ leading to a complete independent set of Cs energy levels based on the findings in this article.

\section{Experimental Method}
\begin{figure}
\centering
\includegraphics[width = 0.95\linewidth]{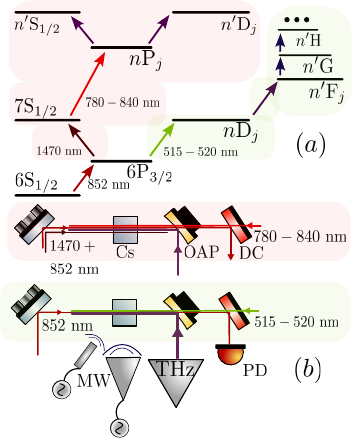}
\caption[Level Diagram and Experimental Layout]{\textbf{Level Diagram and Experimental Layout.} \textbf{(a)} Energy levels of Cs involved in the two-photon (red highlight) and three-photon (green highlight) excitation methods. Once in a Rydberg $np_{j}$ or $nd_{j}$ state, resonant THz and microwave fields are applied to couple to neighbouring states. In the two-photon scheme, further microwave fields can then couple higher angular momentum states.\ \textbf{(b)} Experimental layout of the two- and three-photon schemes. Both employ counter-propagating beams to achieve Doppler-free lineshapes and couple Rydberg states to the ground state via use of EIT. The resulting change in transmission of the probe beam after interaction with the vapor is measured on a photodiode.}
\label{Fig:Exp}
\end{figure}
Throughout this work, Rydberg EIT is used to measure the response of a thermal Cs Vapor to applied fields~\cite{eit}. Resonant laser fields are used to excite atoms to Rydberg states, enabling dipole-allowed transitions to nearby states via resonant THz and/or microwave fields. By monitoring the absorption of the probe (first excitation) laser, the detuning of the THz/microwave field from resonance can be determined, allowing precise measurements of the transition frequency. 
In this work, two different excitation schemes are used to excite Rydberg states in Cs; a two-photon scheme to reach $nd_{j}$ states, and a three-photon scheme for accessing $np_{j}$ states. The energy levels involved in both schemes are shown in Fig.~\ref{Fig:Exp}(a), and the corresponding experimental layouts are shown in Fig.~\ref{Fig:Exp}(b). The initial (probe) laser at 852\,\si{\nano\metre} is common to both schemes and addresses the $6s_{1/2}$, $F = 4$ $\to 6p_{3/2}$, $F^{\prime} = 5$ transition and is stabilised using ground-state polarisation spectroscopy.
In the three-photon scheme (red shading) a second (coupling) laser at 1470\,\si{\nano\metre} co-propagates with the probe and is frequency stabilised to the $6p_{3/2}, F^{\prime} = 5 \to 7s_{1/2}, F^{\prime \prime} = 4$ transition via excited state polarisation spectroscopy~\cite{Carr:12}. To reach Rydberg states a third (Rydberg) laser between 780 - 840\,\si{\nano\metre} counter-propagates with the other two beams. This laser is not frequency stabilised and is instead scanned over the transition of interest. To couple from $np_{j}$ to nearby $ns_{1/2}$ and $nd_{j}$ states, a coherent THz field is incident on the Vapor cell.\par 
The linearly polarized THz beam is emitted from a diagonal horn antenna, collimated by a polytetrafluoroethylene (PTFE) lens, and focused into the cell using an off-axis parabolic (OAP) mirror. The OAP contains a 2~mm through-hole to allow the Rydberg beam to co-propagate with the THz beam. The THz beam is derived from a \textit{Virginia Diodes} amplifier multiplier chain with powers ranging from 5-50 $\upmu$W depending on the frequency emitted and a beam waist of approximately 1.5~mm within the cell. The two THz sources used cover frequencies of $\simeq$ 450 - 800 GHz and 1.01 - 1.06 THz limiting the number of transitions that could be addressed. \par
\begin{figure}
\centering
\includegraphics[width = 0.95\linewidth]{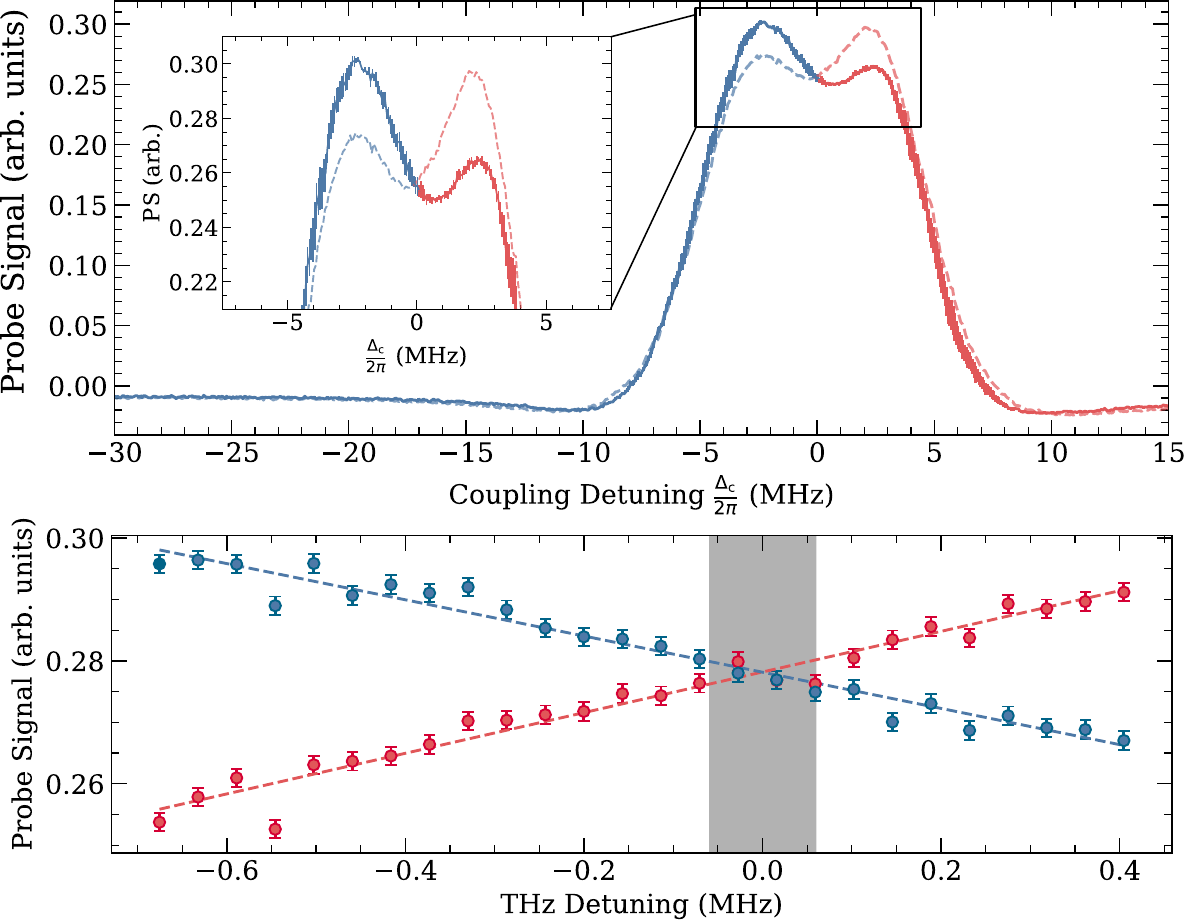}
\caption[Dependence of line shape on THz detuning]{\textbf{Dependence of EIT line shape on THz detuning.} \textit{Top:} Examples of the 2-photon Rydberg EIT signal with an applied THz field that is approximately 0.7\,\si{\mega\hertz} below (solid line) and 0.4 MHz above (dashed line) resonance. When the THz field is resonant with the transition, we expect the height of the two Autler-Townes peaks to be equal. As the THz field is detuned, we see clear asymmetry in the peaks. \textit{Bottom:} Height of the blue detuned (blue) and red-detuned peak (red) as a function of THz frequency - the above peaks corresponding the first and last pair of data points. By fitting straight lines to these points, we can find the resonant transition frequency as the point at which the lines intersect. In this example we measure the frequency of the $18d_{5/2}\rightarrow\,16f_{7/2}$ transition to be $730.40683(6)~\si{\giga\hertz}$, indicated by the shaded region.}
\label{Fig:THz_Detuning}
\end{figure}

In the two-photon scheme (green shading), the second (Rydberg) laser at $515 - 520\,\si{\nano\metre}$ counter-propagates with the probe beam. As in the three-photon case, the Rydberg laser is scanned over the transition of interest. The coherent THz field now couples to $nf_{j}$ states, from which additional applied microwave fields can be used to reach $ng_{j}$ and $nh$ states. Microwave tones are emitted from pyramidal horn antennas for frequencies \textgreater20 GHz and various dipole whip antennas for \textless10 GHz. The microwave fields, estimated to have powers of $\simeq10~\upmu$W within the Vapor cell, are applied perpendicular to the propagation direction of the laser and THz fields. 
In the absence of any applied THz or microwave fields, the transmission of the probe laser exhibits a single Lorentzian EIT peak as the frequency of the Rydberg laser is varied. When THz or microwave fields are applied, Autler-Townes (AT) splitting occurs, dividing the single EIT feature into two Lorentzian peaks, as shown in the upper panel of Fig.\ \ref{Fig:THz_Detuning}. For a resonant applied THz field in the absence of multi-photon couplings~\cite{pritchard2021}, the lineshape is symmetric with equal peak heights. When the field is detuned from resonance asymmetry develops and, for small values of detuning ($\lesssim10\,\si{\mega\hertz}$) the relative peak heights depend linearly on the sign and magnitude of the detuning. By fitting straight lines to the heights of the blue- and red-detuned peaks separately we are able to determine the resonant frequency of the transition by finding the point at which these lines intersect, shown in the lower panel of Fig.~\ref{Fig:THz_Detuning}. For states with $\ell>3$, additional microwave fields introduce couplings between subsequent $\ell$ states, increasing absorption or transmission near the line centre~\cite{allinson2024simultaneous}. These additional fields are similarly detuned to identify transition frequencies.
\section{Quantum Defect Measurements}
\begin{figure*}
\centering
\includegraphics[width = 1\linewidth]{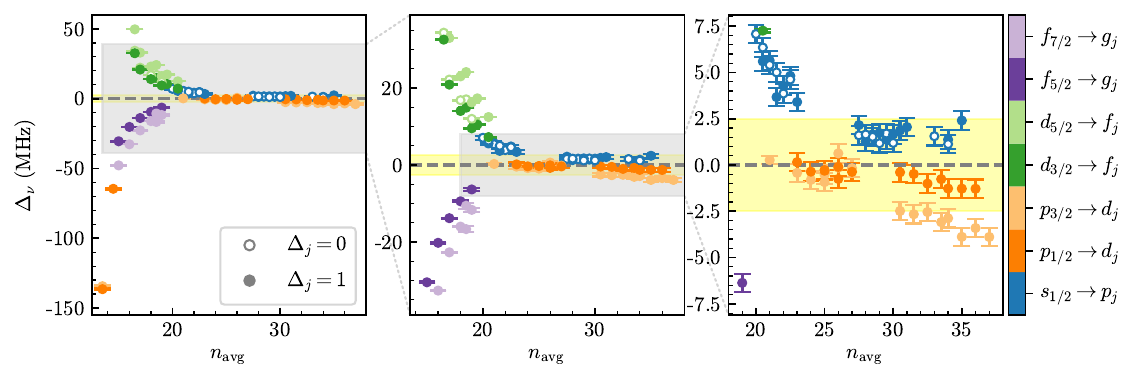}
\caption[Comparison of the quantum defects measured in this work compared to a collation the most recent measurements.]{\textbf{Predicted values of transition frequencies from recent literature compared to our measurements.} The difference ($\Delta_{\nu}$) between our observed transition frequencies and those calculated using quantum defect values from the most recent measurements for each $\ell_{j}$ state. The marker style denotes the change in $j$ with $\Delta_{j} = +1$ (filled) or $\Delta_{j} = 0$ (empty). The colours of the points correspond to the different $\ell$ series considered. Error bars are derived from the determination of the crossing point as shown in Fig.~\ref{Fig:THz_Detuning}. The grey shading indicates the zoom region shown in the subsequent panel. The yellow shading highlight the range in which our optimised values lie.}
\label{fig:Qd_comparison}
\end{figure*}

The frequency of a transition $\nu_0$ between two states described by the quantum numbers $n, \ell, j$ and $n^{\prime}, \ell^{\prime}, j^{\prime}$ is given by
\begin{align}
\nu_0 = & cR_{\infty}\left( \frac{1}{(n - \delta_{\ell, j}(n))^2} - \frac{1}{(n^{\prime} - \delta_{\ell^{\prime},j^{\prime}}(n^{\prime}))^2}\right),
\label{eqn:trans_freq}
\end{align}
where $c$ is the speed of light and $R_{\infty}$ is the Rydberg constant (in $\rm m^{-1}$) for the relevant species (Cs in this case). 
The quantum defects, $\delta_{l,j}(n)$, can be parametrized using the modified Rydberg-Ritz equation as~\cite{martin1980series}
\begin{equation}
\delta_{l,j}(n) = \delta_0 + \frac{\delta_2}{(n - \delta_0)^2} + \frac{\delta_4}{(n - \delta_0)^4} + \frac{\delta_6}{(n - \delta_0)^6} + ... 
\label{eqn:Qd_param}
\end{equation}
where the values for $\delta_{0,2,4,...}$ are coefficients that are distinct for different $\ell$ and $j$. 
Using the most recent reported values of the coefficients $\delta_{0,2,4...}$ (detailed in table~\ref{tab:QDs}) from equation~(\ref{eqn:Qd_param}), the predicted transition frequencies differed significantly from those measured in this work, especially for states where $n<20$. 
Fig.~\ref{fig:Qd_comparison} illustrates the difference between the predicted and measured transition frequencies as a function of the average principal quantum number, $n_{\mathrm{avg}}=(n+n^{\prime})/2$. The plot reveals a distinct structure, where $n$ and $n^{\prime}$ correspond to the initial and final states of the transition, respectively. 
For all series considered, the differences are more apparent at lower $n$. Since the value of the quantum defect $\delta_{\ell, j}$ depends inversely on $n$, we posit that inaccuracies in the underlying empirical coefficients $\delta_{0,2,4}$ are responsible for the observed difference between experiment and theory. These differences are not correlated with frequency, so are not due to any frequency-dependent effects in the terahertz or microwave generation. While there may be slight shifts in our measurements caused by external effects such as for example a dc electric field, any such shifts would typically increase with $n$ and so are unlikely to be the cause of the differences seen here.
The three largest discrepancies are for measurements of the $p_{j}\to d_{j}$ transitions at low $n$ where a difference of over 100\,MHz is observed between our intervals and those predicted by the collated quantum defects. As $n$ increases the observed discrepancy is less apparent, on the order of several MHz at higher $n$, with $np_{3/2}\rightarrow n^{\prime}d_{5/2}$ transitions showing discrepancies of up to 4~MHz. 
The $s_{1/2}\to p_{j}$ series measurements show a systematic discrepancy of several MHz throughout, suggesting some inaccuracy present in the $\delta_0$ coefficient of the $s_{1/2}$, $p_{1/2}$ or $p_{3/2}$ quantum defects. Given that the $p_{j}\to d_{j}$ set of intervals describe the data well at high $n$, and that there is no structure when comparing the $s_{1/2}\to p_{1/2}$ and $s_{1/2}\to p_{3/2}$ intervals, the most likely source of this discrepancy lies in the $s_{1/2}$ quantum defects. 
For the $d_{j}\to f_{j}$ and $f_{j}\to g_{j}$ transitions, discrepancies on the order of tens of MHz are observed with a clear $n$ dependence. 
For the $g_{7/2}$ defect, we observe a constant negative offset implying that the quantum defects from~\cite{Weber:87} overestimate the transition frequencies compared to our measurements.

\begin{figure}
\centering
\includegraphics[width = 1\linewidth]{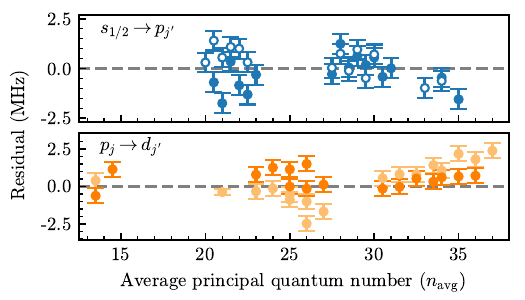}
\includegraphics[width = 1\linewidth]{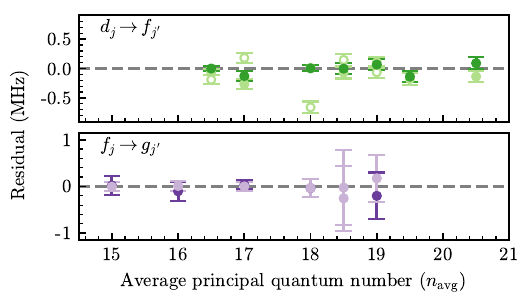}
\caption[Residuals from a global fit to all quantum defects.]{\textbf{Residuals from a global fit to all quantum defects using the transition frequency data.} Measured minus calculated transition frequencies as a function of $n_{\rm{avg}} = (n +n^{\prime})/2$ for optimised values of $\delta_{0,2,4}$. The marker style denotes a transition with initial state $n\ell_{j}$ to a final state $n^{\prime}\ell_{j^{\prime}}^{\prime}$ where $\Delta j = +1$ (filled) or $\Delta j = 0$ (empty). The lighter (darker) points indicate transitions from the higher (lower) fine-structure state, as in Fig~\ref{fig:Qd_comparison}.}
\label{fig:freq_diffs}
\end{figure}

We use our measured transition frequencies to extract measurements of the quantum defect parameters $\delta_{0,2,4}$ in Cs for all states with $\ell\leq 4$ without relying on any data from other sources. 
By using THz frequencies, measuring intervals at significantly lower $n$ than other publications is possible, increasing our sensitivity to changes in quantum defects.
A global fit was performed, optimizing the values of all quantum defects simultaneously. We use equations~(\ref{eqn:trans_freq})~and~(\ref{eqn:Qd_param}) as a theoretical model to calculate the transition frequencies for initial values of $\delta_{0,2,4}$ for all relevant states. A least-squares $\chi^{2}$ minimisation method, implemented via the SciPy \verb|optimise| package in Python, varied the values of $\delta_{0,2,4}$ to minimize the differences between the predicted and measured transition frequencies. 
Table~\ref{tab:QDs} shows our optimised values of the coefficients from equation~(\ref{eqn:Qd_param}) and the residuals of the fit are shown in Fig.~\ref{fig:freq_diffs}. 
While the series expansion in equation~(\ref{eqn:Qd_param}) could be continued to arbitrarily high order, including coefficients higher than $\delta_4$ was not found to significantly improve the minimised value of $\chi^2$ in this work. Therefore the expansion is truncated at fourth order and three parameters ($\delta_{0,2,4}$) are quoted. 
Our values for the parameters $\delta_{0,2,4}$ largely agree with the values reported in~\cite{Weber:87} but, in contrast to the work presented there, do not rely on data from other sources. The error in their measurements is likely underestimated, especially for the higher order quantum defects where values are quoted to a superfluous number of significant figures. While our measurements are of a lower precision than~\cite{DeiglmayrCs}, they agree within error.
\begin{table*}
    \centering
    \renewcommand{\arraystretch}{1.2}
    \resizebox{2\columnwidth}{!}{
    \begin{tabular}{cc |c cc cc cc cc}
            & & $s_{1/2}$ & $p_{1/2}$ & $p_{3/2}$ & $d_{3/2}$ & $d_{5/2}$ & $f_{5/2}$ & $f_{7/2}$ & $g_{7/2}$ & $g_{9/2}$ \\
          \hline
\multirow{3}{*}{This work} & $\delta_{0}$ & 4.049368(3)& 3.591606(3)& 3.559091(3)& 2.475474(4)&  2.466327(4) & 0.033429(3)& 0.033570(3) & 0.007057(5) & 0.007049(4) \\
&$\delta_{2}$ & 0.2333(14)&  0.3569(13) & 0.3657(13)& 0.0051(14)&  0.0101(14)& -0.2025(12)& -0.2016(11)& -0.054(2) & -0.0512(16) \\
&$\delta_{4}$ &  0.8(2)&  0.81(14)& 1.17(14)& -0.06(16)&   0.06(16)& 0.69(13)& 0.45(11) & 0.5(3) & 0.20(19)  \\
\hline
\multirow{5}{*}{Collated Works\footnote{The $s_{1/2}$, $p_{1/2}$, $p_{3/2}$, $d_{5/2}$, quantum defects are from~\cite{DeiglmayrCs}, the $d_{3/2}$ are from~\cite{Lorenzen:84}, the $f_{5/2}$ and $f_{7/2}$ are from~\cite{Fqds} and the $g_{7/2}$ quantum defect is from~\cite{Weber:87}. No $g_{9/2}$ literature value was found.} } & $\delta_{0}$ & 4.0493532(4)& 3.5915871(3) & 3.5590676(3) & 2.4754562&  2.466 3144(6) & 0.03341537(70) &  0.0335646(13) & 0.00703865(70) & - \\
&$\delta_{2}$ & 0.2391(5) &  0.36273(16) & 0.37469(14) & 0.009320 &  0.01381(15)  & -0.2014(16) & -0.2052(29) & -0.049252 & - \\
&$\delta_{4}$ &  0.06(10) &  - & - & -0.43498 & -0.392(12) & - & - & 0.01291 &  - \\
&$\delta_{6}$ &  11(7) &  - & - & -0.76358 &  -1.9(3)& - & - & - & - \\   
&$\delta_{8}$ &  -209(150) &  - & - & -18.0061 &   - & - & - & - & - \\   
    \end{tabular}
    }
    \caption{\textbf{Quantum defect values found in this work compared to a collation the most recent measurements.} Quantum defect coefficients for all spectral series up to $\ell$ = 4, both our optimised values and the most recent literature values. The error in the optimized coefficients was calculated from the covariance matrix generated during the fit. Errors on the literature values are given where available.}
    \label{tab:QDs}
\end{table*}
\par
In the minimisation it was found that there are strong correlations between the optimum values of the $\delta_{0,2,4}$ coefficients for each $\delta_{\ell, j}$.
Whilst this would not affect the optimum value of the fit parameters, it does put bounds on the precision for each coefficient. Additionally, as the fitting minimises transitions between pairs of states, quantum defects for a specific $\ell$ are correlated to those with $\ell\pm1$. Further details of the correlations between parameters can be found in Appendix~\ref{sec:Corrs}.
One of the main use-cases for quantum defect measurements is to calculate the frequency of a specific atomic transition through equation~(\ref{eqn:trans_freq}). In this case the precision of the quantum defect parameters is the dominant source of error and must be accounted for. In the simplest case all parameters ($\delta_{0,2,4}$) can be considered as independent (uncorrelated) and the error in the transition frequency can be calculated by summing the contributions from the uncertainty on each parameter in quadrature~\cite{Hughes:10}. 
Using the optimised values and their precision as quoted in Table~\ref{tab:QDs} we can calculate values of transition frequencies at the MHz level for $n>30$. Incorporating the correlations between fit parameters in the calculation of the error on the predicted transition frequencies is non-trivial but possible, details of two methods are given in Appendix~\ref{sec:Errors}. Doing so increases the precision of transition frequencies calculated using our quoted parameters by up to an order of magnitude. This means that we can predict transition frequencies with a similar precision as those calculated using more precise measurements of $\delta_{0,2,4}$ such as quoted in Table~\ref{tab:QDs}. The entire $27\times27$ covariance matrix derived from our global fit can be found at \cite{data_doi}.
\section{Cs Core Polarizability}
For states with increasing angular momentum, energies can be readily calculated from the polarization energy rather than deriving empirical constants for each distinct $\ell_{j}$ state, as done in the previous section. Here, we use our measured transition frequencies between high-$\ell$ states to determine the dipole ($\alpha_d$) and quadrupole ($\alpha_q$) polarizabilities of the Cs ionic core, Cs${}^{+}$. These constants are used in calculating the energies of states with angular momentum $\ell \geq 5$. This approach provides a systematic framework for determining the quantum defects of all higher $\ell$ states~\cite{Freeman} without their direct measurement.\par
Previous studies of core polarization in alkali metals typically relied on atomic beam experiments, where multi-photon microwave and rf transitions populate excited $n\ell$ states, which were subsequently detected via field ionization~\cite{berl, safinya1980resonance}. Techniques like resonant excitation Stark ionization spectroscopy (RESIS) have observed $\ell$ states as high as 14, though significant DC Stark corrections are required~\cite{Silicon}. Here we use our measured Cs Rydberg high-$\ell$ transitions to extract the dipole and quadrupole polarizabilities of the Cs ionic core.
\par
The energy of a Rydberg state $E_{n\ell j}$ can be expressed as a sum of contributions,
\begin{equation}
\begin{split}
    E_{n\ell j} = & \ E_{I_{\rm Cs}} - \frac{hcR_{\rm Cs}}{n^{2}}\\ & -\Delta E_{\rm pol} - \Delta E_{\rm fs} -\Delta E_{\rm ex} -\Delta E_{\rm pen} - \Delta E_{\rm rel}.
    \end{split}
    \label{eqn:EnAb}
\end{equation}
Here, $E_{I_{\rm Cs}}$ is the ionization energy, and $R_{\rm Cs}$ is the reduced Rydberg constant for Cs. The terms $\Delta E_{\rm pol},\ \Delta E_{\rm fs},\ \Delta E_{\rm ex},\ \Delta E_{\rm pen},\ \Delta E_{\rm rel}$ correspond to polarization energy, fine-structure corrections, exchange energy, penetration energy, and relativistic corrections, respectively~\cite{Sansonetti:81}.
For states with high $\ell$ ($\ell \geq 4$), the penetration and exchange contributions are minimal, so the total energy primarily arises from the polarization energy. The measured energy intervals, $E_{n\ell} - E_{n\ell^{\prime}}$, predominantly reflect differences in polarization energy, which can be expressed in terms of $\alpha_{d}, \alpha_{q}, n $ and $\ell$. The remaining corrections are based on well-known constants. For $\ell \geq 4$, we assume hydrogenic fine-structure and analyze transitions at the centre of mass of the state. Penetration and exchange energies are estimated following the treatment in~\cite{Patil_1994}, which shows good agreement with other calculations of $g$ states~\cite{Sansonetti:81, potas}. While penetration and exchange corrections for $g$ states are small ($\simeq$15 MHz), they are not negligible compared to experimental errors. For $\ell \geq 5$, these corrections are $<$ 100 kHz, well below measurement uncertainty but included for completeness.
\par
We analyze core polarization energies using both adiabatic and non-adiabatic models. The former has historical precedence and enables comparison with prior experimental results, while the latter provides greater accuracy in determination of the Cs$^+$ polarizabilities as it uses a more complete description of the polarization potential~\cite{drachman1982rydberg}.
\subsection{Adiabatic Model}
\begin{figure}
\centering
\hspace*{-0.2cm}\includegraphics[width = \linewidth]{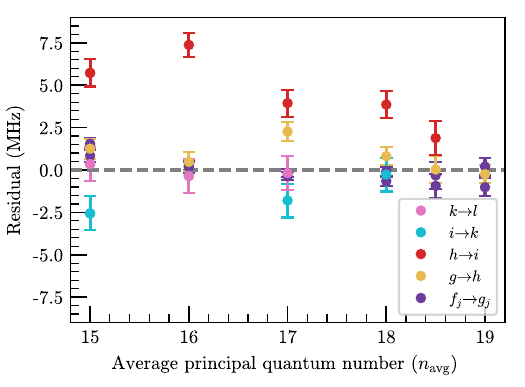}
\caption[Polarisation plot from measurements of the form n$\ell \to$n$\ell^{\prime}$ and global fit to the adiabatic polarisation formula.]{\textbf{Plot of transition frequency intervals obtained from describing $\ell \geq4$ states using an adiabatic model for polarisation energy.} A fit inclusive of  $g_{j}$ states (from the $nf_{j}\to n^{\prime}g_{j}$ intervals) is included with the series of $\ell \geq 4$ measurements taken where the energy of a state is described as a series of contribution from the fine-structure, relativistic, exchange and penetration corrections.}
\label{fig:pol1}
\end{figure}
The adiabatic polarization energy $\Delta E_{\rm pol}$, in atomic units, is given by~\cite{Gallagher:94}
\begin{equation}
    \Delta E_{\rm pol} = -\frac{1}{2}\alpha_{d}^{\prime}\langle r_{n\ell}^{-4}\rangle -\frac{1}{2}\alpha_{q}^{\prime}\langle r_{n\ell}^{-6}\rangle,
    \label{eqn:pol}
\end{equation}
where $\alpha_{d}^{\prime}$ and $\alpha_{q}^{\prime}$ are the effective dipole and quadrupole polarizabilties of the Cs$^{+}$ ion, and $\langle r_{n\ell}^{m}\rangle$ are the radial $m^{\rm th}$ power expectation values of the hydrogenic wavefunction $|n\ell\rangle$~\cite{Bock}. 
For transitions n$\ell$ $\to$ n$\ell^{\prime}$ (of equal $n$) differences in polarization energy can be linearized through
\begin{equation}
   2\frac{\Delta E_{\textrm{pol},n\ell} - \Delta E_{\textrm{pol},n\ell^{\prime}}}{\langle r_{n\ell}^{-4}\rangle - \langle r_{n\ell^{\prime}}^{-4}\rangle} = \alpha_{d}^{\prime} + \alpha_{q}^{\prime}\frac{{\langle r_{n\ell}^{-6}\rangle - \langle r_{n\ell^{\prime}}^{-6}\rangle}}{{\langle r_{n\ell}^{-4}\rangle - \langle r_{n\ell^{\prime}}^{-4}\rangle}},
   \label{eqn:lin}
\end{equation}
where $\Delta E_{\textrm{pol},n\ell} - \Delta E_{\textrm{pol},n\ell^{\prime}}$ are polarisation energies extracted from the measured transitions and the gradient and intercept correspond to the effective quadrupole and dipole polarizability respectively.  
\par
We apply the adiabatic model to two different datasets. One data set uses intervals of the form $n\ell$ $\to$ $n\ell^{\prime}$ where the principal quantum number has not changed. However, given the $nf\to n^{\prime}g$ measurements contain information on the energies of $g$ states, we can neglect the need for identical $n$ intervals and perform a second global fit using all available intervals. In this way, a state $n\ell$ with $\ell \geq 4$ is described by equation ($\ref{eqn:pol}$) instead of unique quantum defects for each $\ell_{j}$ series. The  residuals for the adiabatic model when using all available data is shown in Fig.~\ref{fig:pol1}.\par
The adiabatic model describes the data well and shows good agreement with previous literature values of $\alpha_{d}^{\prime}$ as shown in table~\ref{tab:pol1}. Conversely, the value of $\alpha_{q}^{\prime}$ is in reasonable agreement considering the significant variance between publications.\par
Inspection of the global fit in Fig.~\ref{fig:pol1} suggests that the $nh$ levels are anomalous. This is indicated by slightly elevated $g \to h$ residuals and large $h \to i$ residuals. While the former may be attributed to an overestimation of $E_{\rm pen}$ for the $g$ states, this explanation does not account for the latter measurements, as the penetration energies are on the order of kHz. We considered that this may be the result of a large dc stray electric field, as polarizability scales significantly with $\ell$~\cite{Gallagher:94}. This would decrease the transition frequency for successive $n\ell \to n\ell^{\prime}$ measurements. While this effect is observed in the $h \to i$ dataset, it is not supported by the higher $\ell \to \ell^{\prime}$ transitions, which do not show a corresponding decrease in energy. It is interesting to note that \cite{berl} also observed smaller $h \to i $ intervals than their fit. They made similar observations, but they do not measure as high $\ell$ to further clarify whether this is the result of a stray dc electric field. \par Additionally,~\cite{safinya1980resonance} agrees well with our set of $h\to i$ measurements within their precision. Their interval sizes are not significantly larger, meaning they do not indicate the presence of a stray field in our work. Using a non-adiabatic model as in the following section somewhat resolves this discrepancy. Further measurements of higher $\ell$ states or a larger sample of $n$ may put more confidence on whether this is a systematic error or an unaccounted perturbation.
\begin{table}
    \centering
    \renewcommand{\arraystretch}{1.1}
    \begin{tabular}{cccc}
    \hline
    Reference & $\alpha_{d}^{\prime}$ (a$_{0}^{3}$) & $\alpha_{q}^{\prime}$ (a$_{0}^{5}$) & $\chi^{2}_{\nu,\rm min}$ \\
    \hline 
    This work ($\ell \geq 4)$  & 15.634(17)     &  59.8(13) & 6.0 \\
    This work (global fit) & 15.689(17)     &  55.7(18) & 7.3 \\
    \\
    Safinya \textit{et al.}~\cite{safinya1980resonance} (Expt.)  & 15.544(30)     &  70.7(29) &  \\
    Sansonetti \textit{et al.}~\cite{Sansonetti:81} (Expt.) & 15.79(1) & 38.7(19) &  \\
    Curtis \& Ramanujam~\cite{Curtis:81}\footnote{Using data from \cite{Sansonetti:81,Eriksson_1970,Fredrissin1980}} & 15.759 & 47.990 &  \\
    Weber \& Sansonetti~\cite{Weber:87} (Expt.) & 15.770(3) & 48.9(4) & \\
    \hline 
    \end{tabular}
    \caption{Table of effective Cs$^{+}$ polarizabilities compared to previous experimental results of the effective dipole, $\alpha_{d}^{\prime}$, and quadrupole, $\alpha_{q}^{\prime}$, polarizability. The fitted parameters were found by the minimisation of the reduced $\chi^{2}$ statistic and the errors by re-fitting to $\chi^{2}_{\rm min} + 1$.}
    \label{tab:pol1}
\end{table}
\subsection{Non-Adiabatic Model}
To account for the dynamic interactions between the valence electron and the ionic core, corrections can be applied to equation~(\ref{eqn:pol}) to derive the true static polarizabilities. This work presents the first experimentally determined values of the true static polarizabilities of the Cs${}^{+}$ ion, as all previously reported values have been based on \textit{ab initio} calculations. 
\begin{figure}[h!]
\centering
\hspace*{-0.5cm}\includegraphics[width = 1\linewidth]{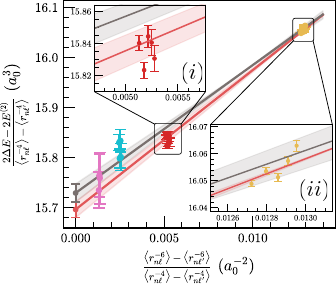}
\caption[Residuals from a non-adiabatic Polarisation Model]{\textbf{Analysis of the data using the non-adiabatic model for two sets of data.} 
Linearization of the data by extending equation (5) to include the non-adiabatic polarization energy from equation (6), allowing extraction of the dipole, $\alpha_{d}$, and quadrupole, $\alpha_{q}$, polarizability. A global fit to all data (including $g_{j}$ states from the $nf_{j}\to n^{\prime}g_{j}$ intervals) is included in the plot (black) for comparison to data where only intervals of the form $n\ell$ $\to$ $n\ell^{\prime}$ (same $n$) are used (red). The shaded regions show the $\pm 1\sigma$ standard errors in the fitted parameters (see table \ref{tab:pol2}). Groups of data points correspond to different sets of transitions with increasing $\ell$: \textbf{(i)} $nh\to ni$ \textbf{(ii)} $ng\to nh$. In the lower left, the transitions $ni\to nk$, $nk\to nl$ are shown, along with the y-intercepts, which correspond to $\alpha_{d}$ for the two fits.}
\label{fig:pol2}
\end{figure}
Two recent treatments addressing adiabatic corrections are  Berl \textit{et al.}~\cite{berl} and Peper \textit{et al.}~\cite{potas}. The former corrects for non-adiabatic effects using the method of Gallagher for measurements in Rb, introducing correcting factors as coefficients in equation~(\ref{eqn:pol}) by considering ionic dipole and quadrupole matrix elements~\cite{Gallagher:94}. The latter follows the treatment of Eissa and \"Opik in $^{39}$K~\cite{Eissa_1967}. However, these analyses were not found to be readily accessible for Cs$^{+}$.
In this paper, we follow the treatment given by Drake and Swainson to extract the true polarisabilities~\cite{DrakeSwan}. The polarisation energy, $\Delta E_{\rm pol}$, can be written as~\cite{Mitroy_2010} 
\begin{equation}
\Delta E_{\rm pol} =  -\frac{1}{2}\alpha_{d}\langle r_{n\ell}^{-4}\rangle- \frac{1}{2}(\alpha_{q}-6\beta)\langle r^{-6}_{n\ell}\rangle + E^{(2)} + ...  
\label{eqn:pol2}
\end{equation}
Here, \(\beta\) is a non-adiabatic correction factor~\cite{DomNA}, estimated via extrapolation of oscillator strength sum rules. Since this extrapolation depends on the dipole polarizability \(\alpha_d\) being extracted, an iterative fitting process is used in which fits are refined until the value of \(\beta\) converges. The term \(E^{(2)}\) represents second-order polarization effects, which are functions of \(\alpha_d\), \(n\), and \(\ell\). Higher-order corrections depend on ionic oscillator strength sums for Cs\(^+\), which are not well characterized.
 \par
The non-adiabatic model describes the data well and an extension of equation (5) to include non-adiabatic corrections is shown in Fig.~\ref{fig:pol2}. We find $\alpha_{d}$ = 15.729(18) $a_{0}^{3}$ and $\alpha_{q}$ = 76.3(1.9) $a_{0}^{5}$ which show good agreement with the theoretical results shown in Table~\ref{tab:pol2}~\cite{JOHNSON1983333,lim2002fully,newCs}, the value of $\alpha_{d}$ we obtain is within $0.5\%$ of these predictions.
However, we note the variance in $\alpha_{q}$ in the literature and the fact that our experimental value is smaller than all theoretical predictions.
The non-adiabatic model somewhat resolves the anomalous $h\to i$ data although there is still some discrepancy when performing a global fit inclusive of the $f\to g$ data. 
\par
Quantum defects for states with $\ell \geq 5$ can then be directly determined from expansions of equation~(\ref{eqn:pol})~\cite{DrakeSwan}, or their energies can be expressed using equation~(\ref{eqn:EnAb}) where their penetration and exchange effects are negligible. The resulting expression is given in Appendix~\ref{sec:expressions}.
\begin{table}
    \centering
    \renewcommand{\arraystretch}{1.1}
    \begin{tabular}{cccc}
    
    \hline
    Reference & $\alpha_{d}$ (a$_{0}^{3}$) & $\alpha_{q}$ (a$_{0}^{5}$) & $\chi^{2}_{\nu,\rm min}$ \\
    \hline 
    This work ($\ell \geq 4$)& 15.696(16) &  78.6(12) & 4.2 \\
    This work (global fit)& 15.729(18)     &  76.3(19) & 5.6 \\
    \\
    Safronova \textit{et al.}~\cite{newCs} (Th.) & 15.84 & - &  \\
    Johnson \textit{et al.}~\cite{JOHNSON1983333} (Th.) & 15.81 & 86.4 &  \\
    Lim \textit{et al.}~\cite{lim2002fully} (Th.) & 15.8(1) & - &  \\
    Sternheimer~\cite{sternheimer1970quadrupole} (Th.) & - & 118.26 &  \\
    Mahan~\cite{mahan1980modified} (Th.) & 15.9 & 108 &  \\
    \hline
    \end{tabular}
    \caption{Table of true Cs$^{+}$ polarisabilities compared to various theoretical results of the dipole, $\alpha_{d}$, and quadrupole, $\alpha_{q}$, polarizability. }
    \label{tab:pol2}
\end{table}
\section{Discussion}
The set of quantum defects presented here for the $s, p, d, f$ and $g$ states and the determination of core polarisabilities will allow more accurate energy levels for reference databases, Rydberg atom interaction potentials and in the calculation of long-range Rydberg molecules~\cite{Weberint,ulr_mol,Eiles_2019}. Appendix~\ref{sec:expressions} details the expressions for estimating energies of $\ell \geq 5$ states using a core Polarizability approach.\par
Since performing the analysis presented in this paper and tabulating the quantum defects listed in Table~\ref{tab:QDs}, more precise measurements for the $s_{1/2}$, $d_{3/2}$ and $d_{5/2}$ states have been made~\cite{Schaffer}. Using these updated values and comparing the resulting theoretical transition frequencies to our measurements did not significantly reduce the differences or remove the structure seen in Fig.~\ref{fig:Qd_comparison}. However the work presented in~\cite{Schaffer} does not include states with $n\leq20$ which is where much of the discrepancy arises.
\par
In previous studies of the core Polarizability of Cs~\cite{Sansonetti:81} and other alkali-metal atoms~\cite{potas,Freeman,Sodium2}, $f_{j}$ has been included as a non-penetrating state. For~\cite{Sansonetti:81}, this gave results for $\alpha_{d}$ and $\alpha_{q}$ that were not in agreement with their $ng_j$ energies. For other species and treatments~\cite{potas,Freeman,Sodium2}, inclusion of penetration and exchange effects for $nf$ states have given good results. This is most likely due to the significantly smaller core penetration experienced by the lighter species, i.e. Na and K. Regardless, the inclusion of our $f$ states in a core polarisation analysis would place better bounds on the value of $\alpha_{q}$ and only slightly change our value of $\alpha_{d}$. We note the importance of including high $\ell$ states beyond $f$ and $g$ states in core polarizability analyses as they allow the deduction of the dipole and quadrupole polarizabilities without reliance on calculation of the exchange and penetration energies.\par 
A data set at a larger range of $n$ would prove better in determining more accurate polarizabilities than extending measurements to higher $\ell$. In our experiment, we are limited both by the range of the THz sources and the two-photon Rydberg laser, which dictate the Rydberg transitions we can measure. A three-photon scheme reaching $f$ Rydberg states would omit the need for a THz source and allow couplings to $g$ states and higher $\ell$ states with microwaves and mm-waves. \par 
A full list of the intervals measured in this work, and their corresponding transition frequencies, is provided in Appendix~\ref{sec:Trans}.

\section{Conclusion}
We have carried out THz and microwave spectroscopy in thermal vapor in order to determine a complete set of quantum defects for Cs and present the first true static polarizability values for the Cs${}^{+}$ ion. Our results improve previous sets of data to allow the estimation of energy levels of Cs to a precision of a few MHz or less. This set of quantum defects will allow a more accurate database for the extensive experimental and theoretical work that relies on such data particularly at lower principal quantum number where transition frequencies lie in the terahertz range~\cite{Wade:17,Downes_2023} \par
The data presented in this article can be found at~\cite{data_doi}.

\section*{Acknowledgements}
The authors would like to thank Matthew Jamieson, Daniel Whiting, Ollie Farley, Ifan Hughes, and Robert Potvliege for fruitful discussion, and Mike Tarbutt for the loan of equipment.\par 
We acknowledge the UK Engineering and Physical Sciences Research Council under grants EP/W033054/1,~EP/V030280/1,~and~EP/W009404/1.

\appendix
\section{Correlation of Fit Parameters}
\label{sec:Corrs}
\begin{figure}
    \centering
    \includegraphics[width=\linewidth]{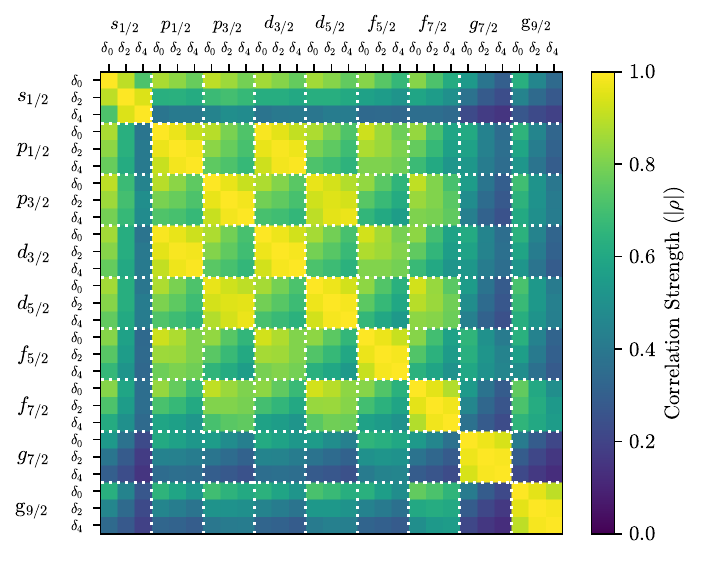}
    \caption{\textbf{Correlation matrix of quantum defects used in the minimised fit.} A color map showing the absolute value of the correlation coefficient $|\rho|$ between pairs of fit parameters. }
    \label{fig:corr}
\end{figure}

When performing the global fit to extract the quantum defects we observe significant correlation between fit parameters which led to larger uncertainties in the reported values. Fig.~\ref{fig:corr} shows the absolute value of the correlation matrix $|\rho|$ where each element is given by
\begin{equation}
    \rho_{ij} = \frac{C_{ij}}{\sqrt{C_{ii}C_{jj}}}
\end{equation}
with $C_{ij}$ being elements of the covariance matrix extracted from the fit \cite{Hughes:10}.
These large correlations are somewhat due to the method used - by measuring intervals between two states, the quantum defects associated with the initial state in the transition are necessarily correlated to the quantum defects that describe the final state. This is seen by inspecting each row in Fig~\ref{fig:corr}. Some level of correlation is seen for each $\ell_{j}$ state across all other states except from the $s_{1/2}$ which shows little. This is most likely attributed to the lowest $n$ interval measured with $s_{1/2}$ is 20 compared to 13, 14 and 15 for the $p, d$ and $f$ states meaning that the fit is not as sensitive to a change in the $s_{1/2}$ quantum defects as the other states. Secondly, the only $\ell$ state accessible from $s_{1/2}$ states are $p_{j}$ states which limits the correlation to other states. Large correlation is seen between the coefficients $\delta_{0,2,4}$ for each electronic state as a change in one can almost entirely be compensated by a change in the other. Lastly, there is notably larger correlation between the $p_{1/2}$ and $d_{3/2}$ quantum defects and similarly between the $p_{3/2}$ and $d_{5/2}$ quantum defects. This is likely due to the large number of measurements between these respective intervals. 

\section{Errors in Transition Frequency Calculations}
\label{sec:Errors}
\begin{figure}[t]
    \centering
    \includegraphics[width=0.9\linewidth]{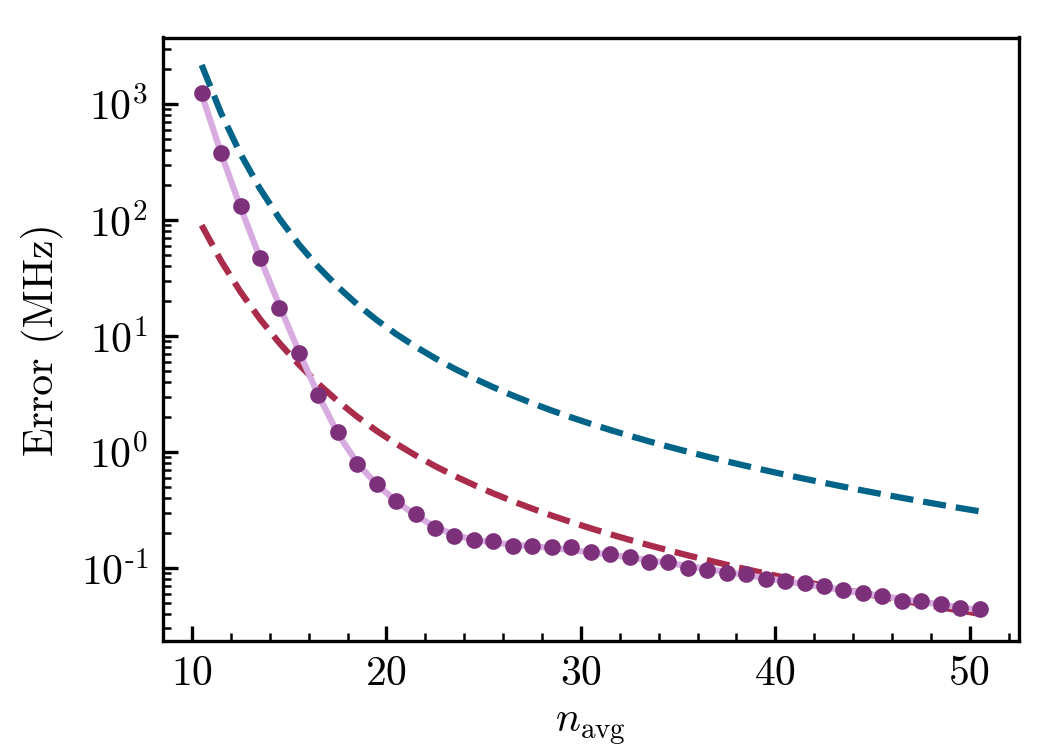}
    \caption{\textbf{Precision of calculated transition frequencies for $\mathbf{np_{3/2}\to(n+1)d_{5/2}}$ transitions} Errors on the transition frequency of $np_{3/2}\to(n+1)d_{5/2}$ transitions as a function of average principal quantum number $n_{\rm{avg}}$. The dashed lines show the result of ignoring covariance for the quantum defects quoted in this work (blue) and recent literature values from table~\ref{tab:1} (red). Including covariances between parameters (purple line \& points) improves the precision of the calculated value, even surpassing the more precise literature for $n\geq16$. The solid line/points represent the analytic Jacobian method (eqn.~\ref{eqn:jac}) and the Monte-Carlo approach (eqn.~\ref{eqn:MC}) respectively.}
    \label{fig:PD_Errs}
\end{figure}

As the optimum parameters for $\delta_{0,2,4}$ are strongly correlated, these correlations need to be taken into account when using the optimized values to predict transition frequencies. Here we will give two methods for doing so. Firstly, taking an analytic approach, the variance in the calculated transition frequency $\sigma^2_{\nu_0}$ as found using equation~(\ref{eqn:trans_freq}) can be expressed as 
\begin{equation}
    \label{eqn:jac}
    \sigma^2_{\nu_0} = \mathbf{JCJ^{T}}
\end{equation}
where 
\begin{equation}
    \mathbf{J} = \left[\frac{\partial \nu_0}{\partial \delta_0}, \frac{\partial \nu_0}{\partial \delta_2},\cdots,\frac{\partial \nu_0}{\partial \delta^{\prime}_2}, \frac{\partial \nu_0}{\partial \delta^{\prime}_4}\right]
\end{equation}
is the Jacobian of equation~(\ref{eqn:trans_freq}) with respect to the parameters $\delta_{0,2,4}$ and $\delta^{\prime}_{0,2,4}$, and $\mathbf{C}$ is the covariance matrix of the form
\begin{equation}
    \mathbf{C} = \begin{bmatrix}
    \sigma^2_{\delta_0} & cov(\delta_0, \delta_2) & \cdots & cov(\delta_0, \delta^{\prime}_4) \\
    cov(\delta_2, \delta_0) & \sigma^2_{\delta_2} & \cdots & cov(\delta_2, \delta^{\prime}_4)\\
    \vdots & \vdots & \ddots & \vdots \\
    cov(\delta^{\prime}_4, \delta_0,) & cov(\delta^{\prime}_4, \delta_2,) & \cdots & \sigma^2_{\delta^{\prime}_4}
    \end{bmatrix}
\end{equation}
with $cov(a, b)$ indicating the covariance between the parameters $a$ and $b$. The 6 terms of the Jacobian can be evaluated analytically but are cumbersome to deal with. 
An alternative approach is to use a Monte-Carlo simulation to estimate the variance of the predicted frequencies. In order to do this, input parameters must have the same covariance as the measured fit parameters. One method to create such a vector of correlated normally distributed random variables $\mathbf{Y}$ is through evaluating
\begin{equation}
    \label{eqn:MC}
    \mathbf{Y} = \mu + \mathbf{\Sigma Z}
\end{equation}
where $\mathbf{Z}$ is a vector of uncorrelated Gaussian random variables and $\mathbf{\Sigma}$ is the Cholesky decomposition of the covariance matrix $\mathbf{C}$ such that $\mathbf{\Sigma \Sigma^T} = \mathbf{C}$. 
Using either of these methods to evaluate the error on frequencies predicted using equations~(\ref{eqn:trans_freq})~and~(\ref{eqn:Qd_param}) results in a significant reduction in the error, meaning that frequencies can be calculated with sub-MHz precision as low as $n=15$. 
An example of the variation in the error of predicted transition frequencies for $np_{3/2}\rightarrow(n+1)d_{5/2}$ transitions is shown in Fig.~\ref{fig:PD_Errs}, and the entire covariance matrix can be found at \cite{data_doi}.
\section{Full Expression for Energy of High \texorpdfstring{$\ell$}{l} States}
\label{sec:expressions}
The energies of states with $\ell \geq 5$, neglecting exchange and penetration contributions, can be expressed as a sum of various components. For completeness, we provide the full expressions in this Appendix. We adopt the adiabatic model for polarization energy, which offers an accurate prediction of energies while maintaining a simple form. The energy $E_{n\ell j}$ is given by:
\begin{equation}
    \begin{split}
    E_{n\ell j} = & \ E_{I_{\rm Cs}} - \frac{hcR_{\rm Cs}}{n^{2}}\\ & -\Delta E_{\rm pol} - \Delta E_{\rm fs} - \Delta E_{\rm rel}
    \end{split}
\end{equation}
where $\Delta E_{\rm pol}$ is given as
\begin{equation}
    \Delta E_{\rm pol} = -hcR_{\rm Cs}\big(\alpha_{d}^{\prime}\langle r_{n\ell}^{-4}\rangle + \alpha_{q}^{\prime}\langle r_{n\ell}^{-6}\rangle\big),
\end{equation}
with $\alpha_{d}^{\prime}$ and $\alpha_{q}^{\prime}$ being the effective dipole and quadrupole polarisabilities found in this paper. The expectation values of the hydrogenic wavefunctions, $\langle r_{n\ell}^{-m}\rangle$, are given as
\begin{equation}
    \langle r_{n\ell}^{-4}\rangle = \frac{3n^{2} - \ell(\ell + 1)}{2n^{5}(\ell - \frac{1}{2})\ell(\ell+\frac{1}{2})(\ell+1)(\ell+\frac{3}{2}) }
\end{equation}
and
\begin{widetext}
\begin{equation}
    \langle r_{n\ell}^{-6}\rangle = \frac{35n^{4} - n^{2}[30\ell(\ell+1)-25]+3(\ell-1)\ell(\ell+1)(\ell+2)}{8n^{7}(\ell-\frac{3}{2})(\ell-1)(\ell-\frac{1}{2})\ell(\ell+\frac{1}{2})(\ell+1)(\ell+\frac{3}{2})(\ell+2)(\ell+\frac{5}{2})}.
\end{equation}
\end{widetext}
The fine-structure correction (spin-orbit coupling), $\Delta E_{\rm fs}$, is given by 
\begin{equation}
    \Delta E_{\rm fs} = -\frac{\alpha^{2}hcR_{\rm Cs}}{2\ell(\ell+\frac{1}{2})(\ell+1)n^{3}}[j(j+1)-\ell(\ell+1) -\frac{3}{4}]
\end{equation}
where $\alpha$ is the fine-structure constant. The relativistic correction, $\Delta E_{\rm rel}$, is given by
\begin{equation}
    \Delta E_{\rm rel} = \frac{\alpha^{2}hcR_{\rm Cs}}{n^3}(\frac{1}{\ell +\frac{1}{2}}-\frac{3}{4n}).
\end{equation}
\section{Table of Transition Frequencies}
\label{sec:Trans}
\begin{table*}
    \centering
    \renewcommand{\arraystretch}{1.05}
    \resizebox{2\columnwidth}{!}{
    \begin{tabular}{cccccc}
    \hline
     \multicolumn{2}{c}{$np$ $\to$ n$^{\prime}$$s$ Measurements} & \multicolumn{2}{c}{$np$ $\to$ n$^{\prime}$$d$ Measurements} & \multicolumn{2}{c}{$nd$ $\to$ n$^{\prime}$$f$ Measurements} \\
        \hline
        Transition& Frequency $\nu$  (GHz) & Transition & Frequency $\nu$  (GHz) & Transition & Frequency $\nu$  (GHz)  \\
         \hline 
        {$20p_{1/2}$ $\to$ $20s_{1/2}$}&-710.9264&{$14p_{1/2}$ $\to$ $13d_{3/2}$}&685.8714&{$18d_{3/2}$ $\to$ $15f_{5/2}$}&-1034.94423(7)\\
        
        {$20p_{1/2}$ $\to$ $21s_{1/2}$}& 770.1388 & {$15p_{3/2}$ $\to$ $14d_{5/2}$}&  548.6153 &{$18d_{3/2}$ $\to$ $16f_{5/2}$}&  746.66404(9)    \\
        {$20p_{3/2}$ $\to$ $21s_{1/2}$}& 721.8937 & {$15p_{1/2}$ $\to$ $14d_{3/2}$}&  519.0775 & {$18d_{5/2}$ $\to$ $15f_{5/2}$}&  -1050.97405(8)\\
        & & & & {$18d_{5/2}$ $\to$ $15f_{7/2}$}&  -1051.24888(9)   \\  

        {$21p_{1/2}$ $\to$ $21s_{1/2}$}& -593.8743 & {$22p_{3/2}$ $\to$ $20d_{5/2}$} & -1025.9229(2) & {$18d_{5/2}$ $\to$ $16f_{5/2}$}&  730.63472(8)   \\
        {$21p_{1/2}$ $\to$ $22s_{1/2}$}& 646.5781 & && {$18d_{5/2}$ $\to$ $16f_{7/2}$}&  730.40683(6)   \\
         {$21p_{3/2}$ $\to$ $22s_{1/2}$}& 606.1677 & {$24p_{1/2}$ $\to$ $22d_{3/2}$}&  -730.6733 &   \\
         {$21p_{3/2}$ $\to$ $21s_{1/2}$}& -634.2831 & {$24p_{3/2}$ $\to$ $22d_{5/2}$}&  -747.7028   & {$19d_{3/2}$ $\to$ $17f_{5/2}$}&  620.57851(8)   \\
        & & & & {$19d_{5/2}$ $\to$ $17f_{5/2}$}&  607.2809(1)   \\     
        {$22p_{1/2}$ $\to$ $22s_{1/2}$}& -501.1722 & {$25p_{1/2}$ $\to$ $23d_{3/2}$}&  -631.0236 & {$19d_{5/2}$ $\to$ $17f_{7/2}$}&  607.09149(5)   \\
        {$22p_{1/2}$ $\to$ $23s_{1/2}$}& 548.1043 & {$25p_{1/2}$ $\to$ $25d_{3/2}$}&  694.2632\\
              {$22p_{3/2}$ $\to$ $23s_{1/2}$}& 513.9218 & {$25p_{3/2}$ $\to$ $23d_{5/2}$}&  -645.8232 & {$20d_{3/2}$ $\to$ $17f_{5/2}$}&  -715.18566(9)   \\
             {$22p_{3/2}$ $\to$ $22s_{1/2}$}& -535.3544 & {$25p_{3/2}$ $\to$ $25d_{5/2}$}&  677.7750& {$20d_{3/2}$ $\to$ $18f_{5/2}$}&  521.36182(9)   \\ 
             & & & & {$20d_{5/2}$ $\to$ $17f_{5/2}$}&  -726.3370(1)   \\ 
            {$23p_{3/2}$ $\to$ $23s_{1/2}$}& -455.9748 & {$26p_{1/2}$ $\to$ $24d_{3/2}$}&  -548.7019 & {$20d_{5/2}$ $\to$ $17f_{7/2}$}&  -726.52729(8)   \\
            & & {$26p_{1/2}$ $\to$ $26d_{3/2}$}&  607.3581 & {$20d_{5/2}$ $\to$ $18f_{5/2}$}&  510.2102(1)   \\  
            {$27p_{1/2}$ $\to$ $29s_{1/2}$}& 719.4515 & {$26p_{3/2}$ $\to$ $24d_{5/2}$}&  -561.6441 & {$20d_{5/2}$ $\to$ $18f_{7/2}$}&  510.0500(1)   \\
            {$27p_{3/2}$ $\to$ $29s_{1/2}$}& 702.8132 & {$26p_{3/2}$ $\to$ $26d_{5/2}$}&  593.0072 \\
             & && & {$21d_{3/2}$ $\to$ $18f_{5/2}$}&  -603.9741(1)   \\
             {$28p_{1/2}$ $\to$ $30s_{1/2}$}& 636.9761 & {$27p_{1/2}$ $\to$ $25d_{3/2}$}&  -480.0988 & {$21d_{5/2}$ $\to$ $18f_{7/2}$}&  -613.5782(1)     \\
             {$28p_{1/2}$ $\to$ $27s_{1/2}$}& -723.7312 &{$27p_{1/2}$ $\to$ $27d_{3/2}$}&  534.3773  \\
        
             {$28p_{3/2}$ $\to$ $30s_{1/2}$}& 622.2992  & {$27p_{3/2}$ $\to$ $25d_{5/2}$}&  -491.4822 & {$22d_{3/2}$ $\to$ $19f_{5/2}$}&  -514.6726(1)   \\
            {$28p_{3/2}$ $\to$ $27s_{1/2}$}& -738.4080 & {$27p_{3/2}$ $\to$ $27d_{5/2}$}&  521.8114 & {$22d_{5/2}$ $\to$ $19f_{7/2}$}&  -522.87675(9)    \\
            & & \\

            {$29p_{1/2}$ $\to$ $31s_{1/2}$}& 566.6445 & {$30p_{1/2}$ $\to$ $31d_{3/2}$}&  674.1317 \\
            {$29p_{1/2}$ $\to$ $28s_{1/2}$}& -639.1752 &{$30p_{3/2}$ $\to$ $31d_{5/2}$}&  665.1313  \\
            {$29p_{3/2}$ $\to$ $31s_{1/2}$}& 553.6318 \\
            {$29p_{3/2}$ $\to$ $28s_{1/2}$}& -652.1882  &{$31p_{3/2}$ $\to$ $32d_{5/2}$}&  597.3845  \\\\
            {$30p_{1/2}$ $\to$ $29s_{1/2}$}& -567.2976 &{$32p_{1/2}$ $\to$ $33d_{3/2}$}&  545.7370 \\
            {$30p_{3/2}$ $\to$ $32s_{1/2}$}& 494.7101 &{$32p_{3/2}$ $\to$ $33d_{5/2}$}&  538.5382  \\
            {$30p_{3/2}$ $\to$ $29s_{1/2}$}& -578.8893  \\
            {$30p_{3/2}$ $\to$ $30s_{1/2}$}& -516.1775 &{$33p_{1/2}$ $\to$ $34d_{3/2}$}&  493.6529  \\
            & &  \\
            {$34p_{1/2}$ $\to$ $32s_{1/2}$}& -653.2059 & {$33p_{1/2}$ $\to$ $35d_{3/2}$}&  694.0853  \\
            & & {$33p_{3/2}$ $\to$ $34d_{5/2}$}&  487.1777 \\
            {$35p_{1/2}$ $\to$ $33s_{1/2}$}& -590.2641 & {$33p_{3/2}$ $\to$ $35d_{5/2}$}&  687.4378   \\
            {$35p_{3/2}$ $\to$ $33s_{1/2}$}& -597.1567 &  \\
            &&{$34p_{1/2}$ $\to$ $36d_{3/2}$}&  630.7592 \\
            {$36p_{3/2}$ $\to$ $34s_{1/2}$}& -541.4330&{$34p_{3/2}$ $\to$ $36d_{5/2}$}&  624.7619\\\\

            & & {$35p_{1/2}$ $\to$ $37d_{3/2}$}&  574.9137\\
            & & {$35p_{3/2}$ $\to$ $37d_{5/2}$}&  569.4835\\\\
            & & {$36p_{3/2}$ $\to$ $38d_{5/2}$}&  520.5438 \\\\
            
           \hline
    \end{tabular}
    }
         \caption{Tabulated values of the $p\to s$, $p\to d$ and $d \to f$ transition frequencies measured using the two- and three-photon method. The $p\to s$ and $p\to d$ measurements have errors of 0.5 MHz. The $d \to f$ measurements have separate errors given for each measurement. A negative sign indicates that the final state lies lower in energy than the initial state.}
    \label{tab:1}
\end{table*}

\begin{table*}
\renewcommand{\arraystretch}{1.05}
    \centering
    \resizebox{2\columnwidth}{!}{
    \begin{tabular}{cc cc cc cc cc}
    \hline
    \multicolumn{2}{c}{$nf\to n^{\prime}g$ ($\ell$ = 4)} & \multicolumn{2}{c}{$ng\to n^{\prime}h$ } & \multicolumn{2}{c}{$nh \to n^{\prime}i$} & \multicolumn{2}{c}{$ni\to n^{\prime}k$} & \multicolumn{2}{c}{$nk\to$ n$^{\prime}l$}  \\
        \hline
          Transition &  $\nu$ / GHz & Transition & $\nu$ / MHz & Transition &  $\nu$ / MHz &Transition &  $\nu$ / MHz& Transition &  $\nu$ / MHz\\
         \hline 
         {$15f_{5/2}$ $\to$ $15g_{7/2}$}& 50.3241(2)   & {$15g_{9/2}$ $\to$ $15h$}&  8700.8(4) &{$15h$ $\to$ $15i$}&  2675.1(5) &$15i$ $\to$ $15k$ &  1016(1) & {$15k$ $\to$ $15l$}&  439(1)\\
        {$15f_{7/2}$ $\to$ $15g_{9/2}$}&  50.6009(1) &    \\   
        \\
    
         {$16f_{5/2}$ $\to$ $16g_{7/2}$}& 41.5844(2)   & {$16g_{9/2}$ $\to$ $16h$}&  7190.5(5) &{$16h$ $\to$ $16i$}&  2206.9(5) & $16i$ $\to$ $16k$&  840(1) & {$16k$ $\to$ $16l$}&  365(1)    \\
            {$16f_{7/2}$ $\to$ $16g_{9/2}$}&  41.8139(1)  &   \\  

            \\
            
            {$17f_{5/2}$ $\to$ $17g_{7/2}$}& 34.7509(1)   & {$17g_{9/2}$ $\to$ $17h$}&  6007.2(5)  & {$17h$ $\to$ $17i$}&  1851.3(4) & {$17i$ $\to$ $17k$}&  705(1)  & {$17k$ $\to$ $17l$}&  306(1)   \\
            {$17f_{7/2}$ $\to$ $17g_{9/2}$}&  34.9430(1)    \\

            \\
            
            {$18f_{5/2}$ $\to$ $18g_{7/2}$}&  29.3321(1) & {$18g_{9/2}$ $\to$ $18h$}&  5071.8(5) & {$18h$ $\to$ $18i$}&  1563.4(5)  & {$18i$ $\to$ $18k$}&  594(1)   \\   
            {$18f_{7/2}$ $\to$ $18g_{9/2}$}&  29.4946(1)  \\\\
            
            {$18f_{7/2}$ $\to$ $19g_{9/2}$}&  1071.33778(7) & {$18g_{9/2}$ $\to$ $19h$} &  1046163.2(3) &{$18h$ $\to$ $19i$}&  1042424.5(4) \\ \\

            {$19f_{5/2}$ $\to$ $19g_{7/2}$}&  24.981(1) & {$19g_{9/2}$ $\to$ $19h$}& 4320.6(5)  \\
           {$19f_{7/2}$ $\to$ $19g_{9/2}$}&  25.1200(8)   \\
           {$19f_{7/2}$ $\to$ $18g_{9/2}$}&  -1016.7236(1) \\
           \hline
    \end{tabular}
    }
    \caption{Tabulated values of the $f\to g$, $g\to h$, $h \to i$, $i \to k$ and $k \to l$ transition frequencies measured using an EIT ladder scheme. The fine structure beyond $f$ is not resolved but we label the $g$ states for completeness and for the dipole allowed transitions. Beyond $g$ we omit the fine structure but assume that the $j = \ell + \frac{1}{2}$ is coupled. A negative sign indicates that the final state lies lower in energy than the initial state.}
    \label{tab:2}
\end{table*}

\section*{References}
\bibliography{All}

\begin{thebibliography}{79}%
\makeatletter
\providecommand \@ifxundefined [1]{%
 \@ifx{#1\undefined}
}%
\providecommand \@ifnum [1]{%
 \ifnum #1\expandafter \@firstoftwo
 \else \expandafter \@secondoftwo
 \fi
}%
\providecommand \@ifx [1]{%
 \ifx #1\expandafter \@firstoftwo
 \else \expandafter \@secondoftwo
 \fi
}%
\providecommand \natexlab [1]{#1}%
\providecommand \enquote  [1]{``#1''}%
\providecommand \bibnamefont  [1]{#1}%
\providecommand \bibfnamefont [1]{#1}%
\providecommand \citenamefont [1]{#1}%
\providecommand \href@noop [0]{\@secondoftwo}%
\providecommand \href [0]{\begingroup \@sanitize@url \@href}%
\providecommand \@href[1]{\@@startlink{#1}\@@href}%
\providecommand \@@href[1]{\endgroup#1\@@endlink}%
\providecommand \@sanitize@url [0]{\catcode `\\12\catcode `\$12\catcode `\&12\catcode `\#12\catcode `\^12\catcode `\_12\catcode `\%12\relax}%
\providecommand \@@startlink[1]{}%
\providecommand \@@endlink[0]{}%
\providecommand \url  [0]{\begingroup\@sanitize@url \@url }%
\providecommand \@url [1]{\endgroup\@href {#1}{\urlprefix }}%
\providecommand \urlprefix  [0]{URL }%
\providecommand \Eprint [0]{\href }%
\providecommand \doibase [0]{http://dx.doi.org/}%
\providecommand \selectlanguage [0]{\@gobble}%
\providecommand \bibinfo  [0]{\@secondoftwo}%
\providecommand \bibfield  [0]{\@secondoftwo}%
\providecommand \translation [1]{[#1]}%
\providecommand \BibitemOpen [0]{}%
\providecommand \bibitemStop [0]{}%
\providecommand \bibitemNoStop [0]{.\EOS\space}%
\providecommand \EOS [0]{\spacefactor3000\relax}%
\providecommand \BibitemShut  [1]{\csname bibitem#1\endcsname}%
\let\auto@bib@innerbib\@empty
\bibitem [{\citenamefont {Safronova}\ \emph {et~al.}(2018)\citenamefont {Safronova}, \citenamefont {Budker}, \citenamefont {DeMille}, \citenamefont {Kimball}, \citenamefont {Derevianko},\ and\ \citenamefont {Clark}}]{newphysics}%
  \BibitemOpen
  \bibfield  {author} {\bibinfo {author} {\bibfnamefont {M.~S.}\ \bibnamefont {Safronova}}, \bibinfo {author} {\bibfnamefont {D.}~\bibnamefont {Budker}}, \bibinfo {author} {\bibfnamefont {D.}~\bibnamefont {DeMille}}, \bibinfo {author} {\bibfnamefont {D.~F.~J.}\ \bibnamefont {Kimball}}, \bibinfo {author} {\bibfnamefont {A.}~\bibnamefont {Derevianko}}, \ and\ \bibinfo {author} {\bibfnamefont {C.~W.}\ \bibnamefont {Clark}},\ }\href {\doibase 10.1103/RevModPhys.90.025008} {\bibfield  {journal} {\bibinfo  {journal} {Rev. Mod. Phys.}\ }\textbf {\bibinfo {volume} {90}},\ \bibinfo {pages} {025008} (\bibinfo {year} {2018})}\BibitemShut {NoStop}%
\bibitem [{\citenamefont {Kozlov}\ \emph {et~al.}(2018)\citenamefont {Kozlov}, \citenamefont {Safronova}, \citenamefont {Crespo L\'opez-Urrutia},\ and\ \citenamefont {Schmidt}}]{Kozlov2018}%
  \BibitemOpen
  \bibfield  {author} {\bibinfo {author} {\bibfnamefont {M.~G.}\ \bibnamefont {Kozlov}}, \bibinfo {author} {\bibfnamefont {M.~S.}\ \bibnamefont {Safronova}}, \bibinfo {author} {\bibfnamefont {J.~R.}\ \bibnamefont {Crespo L\'opez-Urrutia}}, \ and\ \bibinfo {author} {\bibfnamefont {P.~O.}\ \bibnamefont {Schmidt}},\ }\href {\doibase 10.1103/RevModPhys.90.045005} {\bibfield  {journal} {\bibinfo  {journal} {Rev. Mod. Phys.}\ }\textbf {\bibinfo {volume} {90}},\ \bibinfo {pages} {045005} (\bibinfo {year} {2018})}\BibitemShut {NoStop}%
\bibitem [{\citenamefont {Degen}\ \emph {et~al.}(2017)\citenamefont {Degen}, \citenamefont {Reinhard},\ and\ \citenamefont {Cappellaro}}]{qmsensing}%
  \BibitemOpen
  \bibfield  {author} {\bibinfo {author} {\bibfnamefont {C.~L.}\ \bibnamefont {Degen}}, \bibinfo {author} {\bibfnamefont {F.}~\bibnamefont {Reinhard}}, \ and\ \bibinfo {author} {\bibfnamefont {P.}~\bibnamefont {Cappellaro}},\ }\href {\doibase 10.1103/RevModPhys.89.035002} {\bibfield  {journal} {\bibinfo  {journal} {Rev. Mod. Phys.}\ }\textbf {\bibinfo {volume} {89}},\ \bibinfo {pages} {035002} (\bibinfo {year} {2017})}\BibitemShut {NoStop}%
\bibitem [{\citenamefont {Saffman}\ \emph {et~al.}(2010)\citenamefont {Saffman}, \citenamefont {Walker},\ and\ \citenamefont {M\o{}lmer}}]{Saffman2010}%
  \BibitemOpen
  \bibfield  {author} {\bibinfo {author} {\bibfnamefont {M.}~\bibnamefont {Saffman}}, \bibinfo {author} {\bibfnamefont {T.~G.}\ \bibnamefont {Walker}}, \ and\ \bibinfo {author} {\bibfnamefont {K.}~\bibnamefont {M\o{}lmer}},\ }\href {\doibase 10.1103/RevModPhys.82.2313} {\bibfield  {journal} {\bibinfo  {journal} {Rev. Mod. Phys.}\ }\textbf {\bibinfo {volume} {82}},\ \bibinfo {pages} {2313} (\bibinfo {year} {2010})}\BibitemShut {NoStop}%
\bibitem [{\citenamefont {Adams}\ \emph {et~al.}(2019)\citenamefont {Adams}, \citenamefont {Pritchard},\ and\ \citenamefont {Shaffer}}]{Adams_2020}%
  \BibitemOpen
  \bibfield  {author} {\bibinfo {author} {\bibfnamefont {C.~S.}\ \bibnamefont {Adams}}, \bibinfo {author} {\bibfnamefont {J.~D.}\ \bibnamefont {Pritchard}}, \ and\ \bibinfo {author} {\bibfnamefont {J.~P.}\ \bibnamefont {Shaffer}},\ }\href {\doibase 10.1088/1361-6455/ab52ef} {\bibfield  {journal} {\bibinfo  {journal} {Journal of Physics B: Atomic, Molecular and Optical Physics}\ }\textbf {\bibinfo {volume} {53}},\ \bibinfo {pages} {012002} (\bibinfo {year} {2019})}\BibitemShut {NoStop}%
\bibitem [{\citenamefont {Jing}\ \emph {et~al.}(2020)\citenamefont {Jing}, \citenamefont {Hu}, \citenamefont {Ma}, \citenamefont {Zhang}, \citenamefont {Zhang}, \citenamefont {Xiao},\ and\ \citenamefont {Jia}}]{Jing2020}%
  \BibitemOpen
  \bibfield  {author} {\bibinfo {author} {\bibfnamefont {M.}~\bibnamefont {Jing}}, \bibinfo {author} {\bibfnamefont {Y.}~\bibnamefont {Hu}}, \bibinfo {author} {\bibfnamefont {J.}~\bibnamefont {Ma}}, \bibinfo {author} {\bibfnamefont {H.}~\bibnamefont {Zhang}}, \bibinfo {author} {\bibfnamefont {L.}~\bibnamefont {Zhang}}, \bibinfo {author} {\bibfnamefont {L.}~\bibnamefont {Xiao}}, \ and\ \bibinfo {author} {\bibfnamefont {S.}~\bibnamefont {Jia}},\ }\href {\doibase 10.1038/s41567-020-0918-5} {\bibfield  {journal} {\bibinfo  {journal} {Nature Physics}\ }\textbf {\bibinfo {volume} {16}},\ \bibinfo {pages} {911} (\bibinfo {year} {2020})}\BibitemShut {NoStop}%
\bibitem [{\citenamefont {Nowosielski}\ \emph {et~al.}(2024)\citenamefont {Nowosielski}, \citenamefont {Jastrzębski}, \citenamefont {Halavach}, \citenamefont {{\L}ukanowski}, \citenamefont {Jarzyna}, \citenamefont {Mazelanik}, \citenamefont {Wasilewski},\ and\ \citenamefont {Parniak}}]{Nowosielski:24}%
  \BibitemOpen
  \bibfield  {author} {\bibinfo {author} {\bibfnamefont {J.}~\bibnamefont {Nowosielski}}, \bibinfo {author} {\bibfnamefont {M.}~\bibnamefont {Jastrzębski}}, \bibinfo {author} {\bibfnamefont {P.}~\bibnamefont {Halavach}}, \bibinfo {author} {\bibfnamefont {K.}~\bibnamefont {{\L}ukanowski}}, \bibinfo {author} {\bibfnamefont {M.}~\bibnamefont {Jarzyna}}, \bibinfo {author} {\bibfnamefont {M.}~\bibnamefont {Mazelanik}}, \bibinfo {author} {\bibfnamefont {W.}~\bibnamefont {Wasilewski}}, \ and\ \bibinfo {author} {\bibfnamefont {M.}~\bibnamefont {Parniak}},\ }\href {\doibase 10.1364/OE.529977} {\bibfield  {journal} {\bibinfo  {journal} {Opt. Express}\ }\textbf {\bibinfo {volume} {32}},\ \bibinfo {pages} {30027} (\bibinfo {year} {2024})}\BibitemShut {NoStop}%
\bibitem [{\citenamefont {Prajapati}\ \emph {et~al.}(2022)\citenamefont {Prajapati}, \citenamefont {Rotunno}, \citenamefont {Berweger}, \citenamefont {Simons}, \citenamefont {Artusio-Glimpse}, \citenamefont {Voran},\ and\ \citenamefont {Holloway}}]{Nikunjkumap:22}%
  \BibitemOpen
  \bibfield  {author} {\bibinfo {author} {\bibfnamefont {N.}~\bibnamefont {Prajapati}}, \bibinfo {author} {\bibfnamefont {A.~P.}\ \bibnamefont {Rotunno}}, \bibinfo {author} {\bibfnamefont {S.}~\bibnamefont {Berweger}}, \bibinfo {author} {\bibfnamefont {M.~T.}\ \bibnamefont {Simons}}, \bibinfo {author} {\bibfnamefont {A.~B.}\ \bibnamefont {Artusio-Glimpse}}, \bibinfo {author} {\bibfnamefont {S.~D.}\ \bibnamefont {Voran}}, \ and\ \bibinfo {author} {\bibfnamefont {C.~L.}\ \bibnamefont {Holloway}},\ }\href {\doibase 10.1116/5.0098057} {\bibfield  {journal} {\bibinfo  {journal} {AVS Quantum Sci.}\ }\textbf {\bibinfo {volume} {4}},\ \bibinfo {pages} {035001} (\bibinfo {year} {2022})}\BibitemShut {NoStop}%
\bibitem [{\citenamefont {Sedlacek}\ \emph {et~al.}(2012)\citenamefont {Sedlacek}, \citenamefont {Schwettmann}, \citenamefont {K\"{u}bler}, \citenamefont {L\"{o}w}, \citenamefont {Pfau},\ and\ \citenamefont {Shaffer}}]{Sedlacek:12}%
  \BibitemOpen
  \bibfield  {author} {\bibinfo {author} {\bibfnamefont {J.~A.}\ \bibnamefont {Sedlacek}}, \bibinfo {author} {\bibfnamefont {A.}~\bibnamefont {Schwettmann}}, \bibinfo {author} {\bibfnamefont {H.}~\bibnamefont {K\"{u}bler}}, \bibinfo {author} {\bibfnamefont {R.}~\bibnamefont {L\"{o}w}}, \bibinfo {author} {\bibfnamefont {T.}~\bibnamefont {Pfau}}, \ and\ \bibinfo {author} {\bibfnamefont {J.~P.}\ \bibnamefont {Shaffer}},\ }\href@noop {} {\bibfield  {journal} {\bibinfo  {journal} {Nature Physics}\ }\textbf {\bibinfo {volume} {8}},\ \bibinfo {pages} {819} (\bibinfo {year} {2012})}\BibitemShut {NoStop}%
\bibitem [{\citenamefont {Lei}\ and\ \citenamefont {Shi}(2024)}]{Lei:24}%
  \BibitemOpen
  \bibfield  {author} {\bibinfo {author} {\bibfnamefont {M.}~\bibnamefont {Lei}}\ and\ \bibinfo {author} {\bibfnamefont {M.}~\bibnamefont {Shi}},\ }\href {\doibase 10.1364/OL.539090} {\bibfield  {journal} {\bibinfo  {journal} {Opt. Lett.}\ }\textbf {\bibinfo {volume} {49}},\ \bibinfo {pages} {5547} (\bibinfo {year} {2024})}\BibitemShut {NoStop}%
\bibitem [{\citenamefont {Chen}\ \emph {et~al.}(2022)\citenamefont {Chen}, \citenamefont {Reed}, \citenamefont {MacKellar}, \citenamefont {Downes}, \citenamefont {Almuhawish}, \citenamefont {Jamieson}, \citenamefont {Adams},\ and\ \citenamefont {Weatherill}}]{Chen:22}%
  \BibitemOpen
  \bibfield  {author} {\bibinfo {author} {\bibfnamefont {S.}~\bibnamefont {Chen}}, \bibinfo {author} {\bibfnamefont {D.~J.}\ \bibnamefont {Reed}}, \bibinfo {author} {\bibfnamefont {A.~R.}\ \bibnamefont {MacKellar}}, \bibinfo {author} {\bibfnamefont {L.~A.}\ \bibnamefont {Downes}}, \bibinfo {author} {\bibfnamefont {N.~F.~A.}\ \bibnamefont {Almuhawish}}, \bibinfo {author} {\bibfnamefont {M.~J.}\ \bibnamefont {Jamieson}}, \bibinfo {author} {\bibfnamefont {C.~S.}\ \bibnamefont {Adams}}, \ and\ \bibinfo {author} {\bibfnamefont {K.~J.}\ \bibnamefont {Weatherill}},\ }\href {\doibase 10.1364/OPTICA.456761} {\bibfield  {journal} {\bibinfo  {journal} {Optica}\ }\textbf {\bibinfo {volume} {9}},\ \bibinfo {pages} {485} (\bibinfo {year} {2022})}\BibitemShut {NoStop}%
\bibitem [{\citenamefont {Downes}\ \emph {et~al.}(2020)\citenamefont {Downes}, \citenamefont {MacKellar}, \citenamefont {Whiting}, \citenamefont {Bourgenot}, \citenamefont {Adams},\ and\ \citenamefont {Weatherill}}]{Downes:20}%
  \BibitemOpen
  \bibfield  {author} {\bibinfo {author} {\bibfnamefont {L.~A.}\ \bibnamefont {Downes}}, \bibinfo {author} {\bibfnamefont {A.~R.}\ \bibnamefont {MacKellar}}, \bibinfo {author} {\bibfnamefont {D.~J.}\ \bibnamefont {Whiting}}, \bibinfo {author} {\bibfnamefont {C.}~\bibnamefont {Bourgenot}}, \bibinfo {author} {\bibfnamefont {C.~S.}\ \bibnamefont {Adams}}, \ and\ \bibinfo {author} {\bibfnamefont {K.~J.}\ \bibnamefont {Weatherill}},\ }\href {\doibase 10.1103/PhysRevX.10.011027} {\bibfield  {journal} {\bibinfo  {journal} {Phys. Rev. X}\ }\textbf {\bibinfo {volume} {10}},\ \bibinfo {pages} {011027} (\bibinfo {year} {2020})}\BibitemShut {NoStop}%
\bibitem [{\citenamefont {Duspayev}\ \emph {et~al.}(2024)\citenamefont {Duspayev}, \citenamefont {Cardman}, \citenamefont {Anderson},\ and\ \citenamefont {Raithel}}]{highlE}%
  \BibitemOpen
  \bibfield  {author} {\bibinfo {author} {\bibfnamefont {A.}~\bibnamefont {Duspayev}}, \bibinfo {author} {\bibfnamefont {R.}~\bibnamefont {Cardman}}, \bibinfo {author} {\bibfnamefont {D.~A.}\ \bibnamefont {Anderson}}, \ and\ \bibinfo {author} {\bibfnamefont {G.}~\bibnamefont {Raithel}},\ }\href {\doibase 10.1103/PhysRevResearch.6.023138} {\bibfield  {journal} {\bibinfo  {journal} {Phys. Rev. Res.}\ }\textbf {\bibinfo {volume} {6}},\ \bibinfo {pages} {023138} (\bibinfo {year} {2024})}\BibitemShut {NoStop}%
\bibitem [{\citenamefont {Elgee}\ \emph {et~al.}(2023)\citenamefont {Elgee}, \citenamefont {Hill}, \citenamefont {LeBlanc}, \citenamefont {Ko}, \citenamefont {Kunz}, \citenamefont {Meyer},\ and\ \citenamefont {Cox}}]{Meyer:23}%
  \BibitemOpen
  \bibfield  {author} {\bibinfo {author} {\bibfnamefont {P.~K.}\ \bibnamefont {Elgee}}, \bibinfo {author} {\bibfnamefont {J.~C.}\ \bibnamefont {Hill}}, \bibinfo {author} {\bibfnamefont {K.-J.~E.}\ \bibnamefont {LeBlanc}}, \bibinfo {author} {\bibfnamefont {G.~D.}\ \bibnamefont {Ko}}, \bibinfo {author} {\bibfnamefont {P.~D.}\ \bibnamefont {Kunz}}, \bibinfo {author} {\bibfnamefont {D.~H.}\ \bibnamefont {Meyer}}, \ and\ \bibinfo {author} {\bibfnamefont {K.~C.}\ \bibnamefont {Cox}},\ }\href {\doibase 10.1063/5.0158150} {\bibfield  {journal} {\bibinfo  {journal} {Applied Physics Letters}\ }\textbf {\bibinfo {volume} {123}},\ \bibinfo {pages} {084001} (\bibinfo {year} {2023})},\ \Eprint {http://arxiv.org/abs/https://pubs.aip.org/aip/apl/article-pdf/doi/10.1063/5.0158150/18094242/084001\_1\_5.0158150.pdf} {https://pubs.aip.org/aip/apl/article-pdf/doi/10.1063/5.0158150/18094242/084001\_1\_5.0158150.pdf} \BibitemShut {NoStop}%
\bibitem [{\citenamefont {Šibalić}\ \emph {et~al.}(2017)\citenamefont {Šibalić}, \citenamefont {Pritchard}, \citenamefont {Adams},\ and\ \citenamefont {Weatherill}}]{arc1}%
  \BibitemOpen
  \bibfield  {author} {\bibinfo {author} {\bibfnamefont {N.}~\bibnamefont {Šibalić}}, \bibinfo {author} {\bibfnamefont {J.}~\bibnamefont {Pritchard}}, \bibinfo {author} {\bibfnamefont {C.}~\bibnamefont {Adams}}, \ and\ \bibinfo {author} {\bibfnamefont {K.}~\bibnamefont {Weatherill}},\ }\href {\doibase https://doi.org/10.1016/j.cpc.2017.06.015} {\bibfield  {journal} {\bibinfo  {journal} {Computer Physics Communications}\ }\textbf {\bibinfo {volume} {220}},\ \bibinfo {pages} {319} (\bibinfo {year} {2017})}\BibitemShut {NoStop}%
\bibitem [{\citenamefont {Weber}\ and\ \citenamefont {Sansonetti}(1987)}]{Weber:87}%
  \BibitemOpen
  \bibfield  {author} {\bibinfo {author} {\bibfnamefont {K.-H.}\ \bibnamefont {Weber}}\ and\ \bibinfo {author} {\bibfnamefont {C.~J.}\ \bibnamefont {Sansonetti}},\ }\href {\doibase 10.1103/PhysRevA.35.4650} {\bibfield  {journal} {\bibinfo  {journal} {Phys. Rev. A}\ }\textbf {\bibinfo {volume} {35}},\ \bibinfo {pages} {4650} (\bibinfo {year} {1987})}\BibitemShut {NoStop}%
\bibitem [{\citenamefont {Goy}\ \emph {et~al.}(1982)\citenamefont {Goy}, \citenamefont {Raimond}, \citenamefont {Vitrant},\ and\ \citenamefont {Haroche}}]{Goy:82}%
  \BibitemOpen
  \bibfield  {author} {\bibinfo {author} {\bibfnamefont {P.}~\bibnamefont {Goy}}, \bibinfo {author} {\bibfnamefont {J.~M.}\ \bibnamefont {Raimond}}, \bibinfo {author} {\bibfnamefont {G.}~\bibnamefont {Vitrant}}, \ and\ \bibinfo {author} {\bibfnamefont {S.}~\bibnamefont {Haroche}},\ }\href {\doibase 10.1103/PhysRevA.26.2733} {\bibfield  {journal} {\bibinfo  {journal} {Phys. Rev. A}\ }\textbf {\bibinfo {volume} {26}},\ \bibinfo {pages} {2733} (\bibinfo {year} {1982})}\BibitemShut {NoStop}%
\bibitem [{\citenamefont {Sansonetti}\ and\ \citenamefont {Lorenzen}(1984)}]{Sansonetti:84}%
  \BibitemOpen
  \bibfield  {author} {\bibinfo {author} {\bibfnamefont {C.~J.}\ \bibnamefont {Sansonetti}}\ and\ \bibinfo {author} {\bibfnamefont {C.~J.}\ \bibnamefont {Lorenzen}},\ }\href {\doibase 10.1103/PhysRevA.30.1805} {\bibfield  {journal} {\bibinfo  {journal} {Phys. Rev. A}\ }\textbf {\bibinfo {volume} {30}},\ \bibinfo {pages} {1805} (\bibinfo {year} {1984})}\BibitemShut {NoStop}%
\bibitem [{\citenamefont {Bai}\ \emph {et~al.}(2023)\citenamefont {Bai}, \citenamefont {Song}, \citenamefont {Fan}, \citenamefont {Jiao}, \citenamefont {Zhao}, \citenamefont {Jia},\ and\ \citenamefont {Raithel}}]{Fqds}%
  \BibitemOpen
  \bibfield  {author} {\bibinfo {author} {\bibfnamefont {J.}~\bibnamefont {Bai}}, \bibinfo {author} {\bibfnamefont {R.}~\bibnamefont {Song}}, \bibinfo {author} {\bibfnamefont {J.}~\bibnamefont {Fan}}, \bibinfo {author} {\bibfnamefont {Y.}~\bibnamefont {Jiao}}, \bibinfo {author} {\bibfnamefont {J.}~\bibnamefont {Zhao}}, \bibinfo {author} {\bibfnamefont {S.}~\bibnamefont {Jia}}, \ and\ \bibinfo {author} {\bibfnamefont {G.}~\bibnamefont {Raithel}},\ }\href {\doibase 10.1103/PhysRevA.108.022804} {\bibfield  {journal} {\bibinfo  {journal} {Phys. Rev. A}\ }\textbf {\bibinfo {volume} {108}},\ \bibinfo {pages} {022804} (\bibinfo {year} {2023})}\BibitemShut {NoStop}%
\bibitem [{\citenamefont {Deiglmayr}\ \emph {et~al.}(2016)\citenamefont {Deiglmayr}, \citenamefont {Herburger}, \citenamefont {Sa\ss{}mannshausen}, \citenamefont {Jansen}, \citenamefont {Schmutz},\ and\ \citenamefont {Merkt}}]{DeiglmayrCs}%
  \BibitemOpen
  \bibfield  {author} {\bibinfo {author} {\bibfnamefont {J.}~\bibnamefont {Deiglmayr}}, \bibinfo {author} {\bibfnamefont {H.}~\bibnamefont {Herburger}}, \bibinfo {author} {\bibfnamefont {H.}~\bibnamefont {Sa\ss{}mannshausen}}, \bibinfo {author} {\bibfnamefont {P.}~\bibnamefont {Jansen}}, \bibinfo {author} {\bibfnamefont {H.}~\bibnamefont {Schmutz}}, \ and\ \bibinfo {author} {\bibfnamefont {F.}~\bibnamefont {Merkt}},\ }\href {\doibase 10.1103/PhysRevA.93.013424} {\bibfield  {journal} {\bibinfo  {journal} {Phys. Rev. A}\ }\textbf {\bibinfo {volume} {93}},\ \bibinfo {pages} {013424} (\bibinfo {year} {2016})}\BibitemShut {NoStop}%
\bibitem [{\citenamefont {Sa\ss{}mannshausen}\ \emph {et~al.}(2013)\citenamefont {Sa\ss{}mannshausen}, \citenamefont {Merkt},\ and\ \citenamefont {Deiglmayr}}]{DeiglmayrCs2}%
  \BibitemOpen
  \bibfield  {author} {\bibinfo {author} {\bibfnamefont {H.}~\bibnamefont {Sa\ss{}mannshausen}}, \bibinfo {author} {\bibfnamefont {F.}~\bibnamefont {Merkt}}, \ and\ \bibinfo {author} {\bibfnamefont {J.}~\bibnamefont {Deiglmayr}},\ }\href {\doibase 10.1103/PhysRevA.87.032519} {\bibfield  {journal} {\bibinfo  {journal} {Phys. Rev. A}\ }\textbf {\bibinfo {volume} {87}},\ \bibinfo {pages} {032519} (\bibinfo {year} {2013})}\BibitemShut {NoStop}%
\bibitem [{\citenamefont {Lorenzen}\ and\ \citenamefont {Niemax}(1984)}]{Lorenzen:84}%
  \BibitemOpen
  \bibfield  {author} {\bibinfo {author} {\bibfnamefont {C.-J.}\ \bibnamefont {Lorenzen}}\ and\ \bibinfo {author} {\bibfnamefont {K.}~\bibnamefont {Niemax}},\ }\href {\doibase 10.1007/BF01419370} {\bibfield  {journal} {\bibinfo  {journal} {Zeitschrift f{\"u}r Physik A Atoms and Nuclei}\ }\textbf {\bibinfo {volume} {315}},\ \bibinfo {pages} {127} (\bibinfo {year} {1984})}\BibitemShut {NoStop}%
\bibitem [{\citenamefont {Freeman}\ and\ \citenamefont {Kleppner}(1976)}]{Freeman}%
  \BibitemOpen
  \bibfield  {author} {\bibinfo {author} {\bibfnamefont {R.~R.}\ \bibnamefont {Freeman}}\ and\ \bibinfo {author} {\bibfnamefont {D.}~\bibnamefont {Kleppner}},\ }\href {\doibase 10.1103/PhysRevA.14.1614} {\bibfield  {journal} {\bibinfo  {journal} {Phys. Rev. A}\ }\textbf {\bibinfo {volume} {14}},\ \bibinfo {pages} {1614} (\bibinfo {year} {1976})}\BibitemShut {NoStop}%
\bibitem [{\citenamefont {Lundeen}(2005)}]{LUNDEEN2005161}%
  \BibitemOpen
  \bibfield  {author} {\bibinfo {author} {\bibfnamefont {S.~R.}\ \bibnamefont {Lundeen}}\ }(\bibinfo  {publisher} {Academic Press},\ \bibinfo {year} {2005})\ pp.\ \bibinfo {pages} {161--208}\BibitemShut {NoStop}%
\bibitem [{\citenamefont {Cooke}\ \emph {et~al.}(1977)\citenamefont {Cooke}, \citenamefont {Gallagher}, \citenamefont {Hill},\ and\ \citenamefont {Edelstein}}]{lithium}%
  \BibitemOpen
  \bibfield  {author} {\bibinfo {author} {\bibfnamefont {W.~E.}\ \bibnamefont {Cooke}}, \bibinfo {author} {\bibfnamefont {T.~F.}\ \bibnamefont {Gallagher}}, \bibinfo {author} {\bibfnamefont {R.~M.}\ \bibnamefont {Hill}}, \ and\ \bibinfo {author} {\bibfnamefont {S.~A.}\ \bibnamefont {Edelstein}},\ }\href {\doibase 10.1103/PhysRevA.16.1141} {\bibfield  {journal} {\bibinfo  {journal} {Phys. Rev. A}\ }\textbf {\bibinfo {volume} {16}},\ \bibinfo {pages} {1141} (\bibinfo {year} {1977})}\BibitemShut {NoStop}%
\bibitem [{\citenamefont {Gray}\ \emph {et~al.}(1988{\natexlab{a}})\citenamefont {Gray}, \citenamefont {Sun},\ and\ \citenamefont {MacAdam}}]{sodium}%
  \BibitemOpen
  \bibfield  {author} {\bibinfo {author} {\bibfnamefont {L.~G.}\ \bibnamefont {Gray}}, \bibinfo {author} {\bibfnamefont {X.}~\bibnamefont {Sun}}, \ and\ \bibinfo {author} {\bibfnamefont {K.~B.}\ \bibnamefont {MacAdam}},\ }\href {\doibase 10.1103/PhysRevA.38.4985} {\bibfield  {journal} {\bibinfo  {journal} {Phys. Rev. A}\ }\textbf {\bibinfo {volume} {38}},\ \bibinfo {pages} {4985} (\bibinfo {year} {1988}{\natexlab{a}})}\BibitemShut {NoStop}%
\bibitem [{\citenamefont {Peper}\ \emph {et~al.}(2019)\citenamefont {Peper}, \citenamefont {Helmrich}, \citenamefont {Butscher}, \citenamefont {Agner}, \citenamefont {Schmutz}, \citenamefont {Merkt},\ and\ \citenamefont {Deiglmayr}}]{potas}%
  \BibitemOpen
  \bibfield  {author} {\bibinfo {author} {\bibfnamefont {M.}~\bibnamefont {Peper}}, \bibinfo {author} {\bibfnamefont {F.}~\bibnamefont {Helmrich}}, \bibinfo {author} {\bibfnamefont {J.}~\bibnamefont {Butscher}}, \bibinfo {author} {\bibfnamefont {J.~A.}\ \bibnamefont {Agner}}, \bibinfo {author} {\bibfnamefont {H.}~\bibnamefont {Schmutz}}, \bibinfo {author} {\bibfnamefont {F.}~\bibnamefont {Merkt}}, \ and\ \bibinfo {author} {\bibfnamefont {J.}~\bibnamefont {Deiglmayr}},\ }\href {\doibase 10.1103/PhysRevA.100.012501} {\bibfield  {journal} {\bibinfo  {journal} {Phys. Rev. A}\ }\textbf {\bibinfo {volume} {100}},\ \bibinfo {pages} {012501} (\bibinfo {year} {2019})}\BibitemShut {NoStop}%
\bibitem [{\citenamefont {Komara}\ \emph {et~al.}(2005)\citenamefont {Komara}, \citenamefont {Gearba}, \citenamefont {Fehrenbach},\ and\ \citenamefont {Lundeen}}]{Komara_2005}%
  \BibitemOpen
  \bibfield  {author} {\bibinfo {author} {\bibfnamefont {R.~A.}\ \bibnamefont {Komara}}, \bibinfo {author} {\bibfnamefont {M.~A.}\ \bibnamefont {Gearba}}, \bibinfo {author} {\bibfnamefont {C.~W.}\ \bibnamefont {Fehrenbach}}, \ and\ \bibinfo {author} {\bibfnamefont {S.~R.}\ \bibnamefont {Lundeen}},\ }\href {\doibase 10.1088/0953-4075/38/2/007} {\bibfield  {journal} {\bibinfo  {journal} {Journal of Physics B: Atomic, Molecular and Optical Physics}\ }\textbf {\bibinfo {volume} {38}},\ \bibinfo {pages} {S87} (\bibinfo {year} {2005})}\BibitemShut {NoStop}%
\bibitem [{\citenamefont {Snow}\ \emph {et~al.}(2005)\citenamefont {Snow}, \citenamefont {Gearba}, \citenamefont {Komara}, \citenamefont {Lundeen},\ and\ \citenamefont {Sturrus}}]{barium}%
  \BibitemOpen
  \bibfield  {author} {\bibinfo {author} {\bibfnamefont {E.~L.}\ \bibnamefont {Snow}}, \bibinfo {author} {\bibfnamefont {M.~A.}\ \bibnamefont {Gearba}}, \bibinfo {author} {\bibfnamefont {R.~A.}\ \bibnamefont {Komara}}, \bibinfo {author} {\bibfnamefont {S.~R.}\ \bibnamefont {Lundeen}}, \ and\ \bibinfo {author} {\bibfnamefont {W.~G.}\ \bibnamefont {Sturrus}},\ }\href {\doibase 10.1103/PhysRevA.71.022510} {\bibfield  {journal} {\bibinfo  {journal} {Phys. Rev. A}\ }\textbf {\bibinfo {volume} {71}},\ \bibinfo {pages} {022510} (\bibinfo {year} {2005})}\BibitemShut {NoStop}%
\bibitem [{\citenamefont {Jacobson}\ \emph {et~al.}(1996)\citenamefont {Jacobson}, \citenamefont {Labelle}, \citenamefont {Sturrus}, \citenamefont {Ward},\ and\ \citenamefont {Lundeen}}]{nitrogen}%
  \BibitemOpen
  \bibfield  {author} {\bibinfo {author} {\bibfnamefont {P.~L.}\ \bibnamefont {Jacobson}}, \bibinfo {author} {\bibfnamefont {R.~D.}\ \bibnamefont {Labelle}}, \bibinfo {author} {\bibfnamefont {W.~G.}\ \bibnamefont {Sturrus}}, \bibinfo {author} {\bibfnamefont {R.~F.}\ \bibnamefont {Ward}}, \ and\ \bibinfo {author} {\bibfnamefont {S.~R.}\ \bibnamefont {Lundeen}},\ }\href {\doibase 10.1103/PhysRevA.54.314} {\bibfield  {journal} {\bibinfo  {journal} {Phys. Rev. A}\ }\textbf {\bibinfo {volume} {54}},\ \bibinfo {pages} {314} (\bibinfo {year} {1996})}\BibitemShut {NoStop}%
\bibitem [{\citenamefont {Deck}\ \emph {et~al.}(1993)\citenamefont {Deck}, \citenamefont {Hessels},\ and\ \citenamefont {Lundeen}}]{sulfur}%
  \BibitemOpen
  \bibfield  {author} {\bibinfo {author} {\bibfnamefont {F.~J.}\ \bibnamefont {Deck}}, \bibinfo {author} {\bibfnamefont {E.~A.}\ \bibnamefont {Hessels}}, \ and\ \bibinfo {author} {\bibfnamefont {S.~R.}\ \bibnamefont {Lundeen}},\ }\href {\doibase 10.1103/PhysRevA.48.4400} {\bibfield  {journal} {\bibinfo  {journal} {Phys. Rev. A}\ }\textbf {\bibinfo {volume} {48}},\ \bibinfo {pages} {4400} (\bibinfo {year} {1993})}\BibitemShut {NoStop}%
\bibitem [{\citenamefont {Drachman}(1982{\natexlab{a}})}]{PhysRevA.26.1228}%
  \BibitemOpen
  \bibfield  {author} {\bibinfo {author} {\bibfnamefont {R.~J.}\ \bibnamefont {Drachman}},\ }\href {\doibase 10.1103/PhysRevA.26.1228} {\bibfield  {journal} {\bibinfo  {journal} {Phys. Rev. A}\ }\textbf {\bibinfo {volume} {26}},\ \bibinfo {pages} {1228} (\bibinfo {year} {1982}{\natexlab{a}})}\BibitemShut {NoStop}%
\bibitem [{\citenamefont {Jacobson}\ \emph {et~al.}(1997)\citenamefont {Jacobson}, \citenamefont {Fisher}, \citenamefont {Fehrenbach}, \citenamefont {Sturrus},\ and\ \citenamefont {Lundeen}}]{PhysRevA.56.R4361}%
  \BibitemOpen
  \bibfield  {author} {\bibinfo {author} {\bibfnamefont {P.~L.}\ \bibnamefont {Jacobson}}, \bibinfo {author} {\bibfnamefont {D.~S.}\ \bibnamefont {Fisher}}, \bibinfo {author} {\bibfnamefont {C.~W.}\ \bibnamefont {Fehrenbach}}, \bibinfo {author} {\bibfnamefont {W.~G.}\ \bibnamefont {Sturrus}}, \ and\ \bibinfo {author} {\bibfnamefont {S.~R.}\ \bibnamefont {Lundeen}},\ }\href {\doibase 10.1103/PhysRevA.56.R4361} {\bibfield  {journal} {\bibinfo  {journal} {Phys. Rev. A}\ }\textbf {\bibinfo {volume} {56}},\ \bibinfo {pages} {R4361} (\bibinfo {year} {1997})}\BibitemShut {NoStop}%
\bibitem [{\citenamefont {Mitroy}\ \emph {et~al.}(2010)\citenamefont {Mitroy}, \citenamefont {Safronova},\ and\ \citenamefont {Clark}}]{Mitroy_2010}%
  \BibitemOpen
  \bibfield  {author} {\bibinfo {author} {\bibfnamefont {J.}~\bibnamefont {Mitroy}}, \bibinfo {author} {\bibfnamefont {M.~S.}\ \bibnamefont {Safronova}}, \ and\ \bibinfo {author} {\bibfnamefont {C.~W.}\ \bibnamefont {Clark}},\ }\href {\doibase 10.1088/0953-4075/43/20/202001} {\bibfield  {journal} {\bibinfo  {journal} {Journal of Physics B: Atomic, Molecular and Optical Physics}\ }\textbf {\bibinfo {volume} {43}},\ \bibinfo {pages} {202001} (\bibinfo {year} {2010})}\BibitemShut {NoStop}%
\bibitem [{\citenamefont {Schlagm\"uller}\ \emph {et~al.}(2016)\citenamefont {Schlagm\"uller}, \citenamefont {Liebisch}, \citenamefont {Engel}, \citenamefont {Kleinbach}, \citenamefont {B\"ottcher}, \citenamefont {Hermann}, \citenamefont {Westphal}, \citenamefont {Gaj}, \citenamefont {L\"ow}, \citenamefont {Hofferberth}, \citenamefont {Pfau}, \citenamefont {P\'erez-R\'{\i}os},\ and\ \citenamefont {Greene}}]{core_int_1}%
  \BibitemOpen
  \bibfield  {author} {\bibinfo {author} {\bibfnamefont {M.}~\bibnamefont {Schlagm\"uller}}, \bibinfo {author} {\bibfnamefont {T.~C.}\ \bibnamefont {Liebisch}}, \bibinfo {author} {\bibfnamefont {F.}~\bibnamefont {Engel}}, \bibinfo {author} {\bibfnamefont {K.~S.}\ \bibnamefont {Kleinbach}}, \bibinfo {author} {\bibfnamefont {F.}~\bibnamefont {B\"ottcher}}, \bibinfo {author} {\bibfnamefont {U.}~\bibnamefont {Hermann}}, \bibinfo {author} {\bibfnamefont {K.~M.}\ \bibnamefont {Westphal}}, \bibinfo {author} {\bibfnamefont {A.}~\bibnamefont {Gaj}}, \bibinfo {author} {\bibfnamefont {R.}~\bibnamefont {L\"ow}}, \bibinfo {author} {\bibfnamefont {S.}~\bibnamefont {Hofferberth}}, \bibinfo {author} {\bibfnamefont {T.}~\bibnamefont {Pfau}}, \bibinfo {author} {\bibfnamefont {J.}~\bibnamefont {P\'erez-R\'{\i}os}}, \ and\ \bibinfo {author} {\bibfnamefont {C.~H.}\ \bibnamefont {Greene}},\ }\href {\doibase 10.1103/PhysRevX.6.031020} {\bibfield  {journal} {\bibinfo  {journal} {Phys. Rev. X}\ }\textbf {\bibinfo {volume} {6}},\
  \bibinfo {pages} {031020} (\bibinfo {year} {2016})}\BibitemShut {NoStop}%
\bibitem [{\citenamefont {Allmendinger}\ \emph {et~al.}(2016)\citenamefont {Allmendinger}, \citenamefont {Deiglmayr}, \citenamefont {Höveler}, \citenamefont {Schullian},\ and\ \citenamefont {Merkt}}]{core_int_2}%
  \BibitemOpen
  \bibfield  {author} {\bibinfo {author} {\bibfnamefont {P.}~\bibnamefont {Allmendinger}}, \bibinfo {author} {\bibfnamefont {J.}~\bibnamefont {Deiglmayr}}, \bibinfo {author} {\bibfnamefont {K.}~\bibnamefont {Höveler}}, \bibinfo {author} {\bibfnamefont {O.}~\bibnamefont {Schullian}}, \ and\ \bibinfo {author} {\bibfnamefont {F.}~\bibnamefont {Merkt}},\ }\href {\doibase 10.1063/1.4972130} {\bibfield  {journal} {\bibinfo  {journal} {The Journal of Chemical Physics}\ }\textbf {\bibinfo {volume} {145}},\ \bibinfo {pages} {244316} (\bibinfo {year} {2016})},\ \Eprint {http://arxiv.org/abs/https://pubs.aip.org/aip/jcp/article-pdf/doi/10.1063/1.4972130/15522196/244316\_1\_online.pdf} {https://pubs.aip.org/aip/jcp/article-pdf/doi/10.1063/1.4972130/15522196/244316\_1\_online.pdf} \BibitemShut {NoStop}%
\bibitem [{\citenamefont {Marinescu}\ \emph {et~al.}(1994)\citenamefont {Marinescu}, \citenamefont {Sadeghpour},\ and\ \citenamefont {Dalgarno}}]{atom_atom}%
  \BibitemOpen
  \bibfield  {author} {\bibinfo {author} {\bibfnamefont {M.}~\bibnamefont {Marinescu}}, \bibinfo {author} {\bibfnamefont {H.~R.}\ \bibnamefont {Sadeghpour}}, \ and\ \bibinfo {author} {\bibfnamefont {A.}~\bibnamefont {Dalgarno}},\ }\href {\doibase 10.1103/PhysRevA.49.982} {\bibfield  {journal} {\bibinfo  {journal} {Phys. Rev. A}\ }\textbf {\bibinfo {volume} {49}},\ \bibinfo {pages} {982} (\bibinfo {year} {1994})}\BibitemShut {NoStop}%
\bibitem [{\citenamefont {Westergaard}\ \emph {et~al.}(2011)\citenamefont {Westergaard}, \citenamefont {Lodewyck}, \citenamefont {Lorini}, \citenamefont {Lecallier}, \citenamefont {Burt}, \citenamefont {Zawada}, \citenamefont {Millo},\ and\ \citenamefont {Lemonde}}]{StarkPol}%
  \BibitemOpen
  \bibfield  {author} {\bibinfo {author} {\bibfnamefont {P.~G.}\ \bibnamefont {Westergaard}}, \bibinfo {author} {\bibfnamefont {J.}~\bibnamefont {Lodewyck}}, \bibinfo {author} {\bibfnamefont {L.}~\bibnamefont {Lorini}}, \bibinfo {author} {\bibfnamefont {A.}~\bibnamefont {Lecallier}}, \bibinfo {author} {\bibfnamefont {E.~A.}\ \bibnamefont {Burt}}, \bibinfo {author} {\bibfnamefont {M.}~\bibnamefont {Zawada}}, \bibinfo {author} {\bibfnamefont {J.}~\bibnamefont {Millo}}, \ and\ \bibinfo {author} {\bibfnamefont {P.}~\bibnamefont {Lemonde}},\ }\href {\doibase 10.1103/PhysRevLett.106.210801} {\bibfield  {journal} {\bibinfo  {journal} {Phys. Rev. Lett.}\ }\textbf {\bibinfo {volume} {106}},\ \bibinfo {pages} {210801} (\bibinfo {year} {2011})}\BibitemShut {NoStop}%
\bibitem [{\citenamefont {Safronova}\ \emph {et~al.}(2012)\citenamefont {Safronova}, \citenamefont {Kozlov},\ and\ \citenamefont {Clark}}]{bbratom}%
  \BibitemOpen
  \bibfield  {author} {\bibinfo {author} {\bibfnamefont {M.~S.}\ \bibnamefont {Safronova}}, \bibinfo {author} {\bibfnamefont {M.~G.}\ \bibnamefont {Kozlov}}, \ and\ \bibinfo {author} {\bibfnamefont {C.~W.}\ \bibnamefont {Clark}},\ }\href {\doibase 10.1109/TUFFC.2012.2213} {\bibfield  {journal} {\bibinfo  {journal} {IEEE Transactions on Ultrasonics, Ferroelectrics, and Frequency Control}\ }\textbf {\bibinfo {volume} {59}},\ \bibinfo {pages} {439} (\bibinfo {year} {2012})}\BibitemShut {NoStop}%
\bibitem [{\citenamefont {Cohen}\ and\ \citenamefont {Thompson}(2021)}]{qc_circ}%
  \BibitemOpen
  \bibfield  {author} {\bibinfo {author} {\bibfnamefont {S.~R.}\ \bibnamefont {Cohen}}\ and\ \bibinfo {author} {\bibfnamefont {J.~D.}\ \bibnamefont {Thompson}},\ }\href {\doibase 10.1103/PRXQuantum.2.030322} {\bibfield  {journal} {\bibinfo  {journal} {PRX Quantum}\ }\textbf {\bibinfo {volume} {2}},\ \bibinfo {pages} {030322} (\bibinfo {year} {2021})}\BibitemShut {NoStop}%
\bibitem [{\citenamefont {Hare}\ \emph {et~al.}(1993)\citenamefont {Hare}, \citenamefont {Nussenzweig}, \citenamefont {Gabbanini}, \citenamefont {Weidemuller}, \citenamefont {Goy}, \citenamefont {Gross},\ and\ \citenamefont {Haroche}}]{circ1}%
  \BibitemOpen
  \bibfield  {author} {\bibinfo {author} {\bibfnamefont {J.}~\bibnamefont {Hare}}, \bibinfo {author} {\bibfnamefont {A.}~\bibnamefont {Nussenzweig}}, \bibinfo {author} {\bibfnamefont {C.}~\bibnamefont {Gabbanini}}, \bibinfo {author} {\bibfnamefont {M.}~\bibnamefont {Weidemuller}}, \bibinfo {author} {\bibfnamefont {P.}~\bibnamefont {Goy}}, \bibinfo {author} {\bibfnamefont {M.}~\bibnamefont {Gross}}, \ and\ \bibinfo {author} {\bibfnamefont {S.}~\bibnamefont {Haroche}},\ }\href {\doibase 10.1109/19.278576} {\bibfield  {journal} {\bibinfo  {journal} {IEEE Transactions on Instrumentation and Measurement}\ }\textbf {\bibinfo {volume} {42}},\ \bibinfo {pages} {331} (\bibinfo {year} {1993})}\BibitemShut {NoStop}%
\bibitem [{\citenamefont {Jentschura}\ \emph {et~al.}(2010)\citenamefont {Jentschura}, \citenamefont {Mohr},\ and\ \citenamefont {Tan}}]{circ2}%
  \BibitemOpen
  \bibfield  {author} {\bibinfo {author} {\bibfnamefont {U.~D.}\ \bibnamefont {Jentschura}}, \bibinfo {author} {\bibfnamefont {P.~J.}\ \bibnamefont {Mohr}}, \ and\ \bibinfo {author} {\bibfnamefont {J.~N.}\ \bibnamefont {Tan}},\ }\href {\doibase 10.1088/0953-4075/43/7/074002} {\bibfield  {journal} {\bibinfo  {journal} {Journal of Physics B: Atomic, Molecular and Optical Physics}\ }\textbf {\bibinfo {volume} {43}},\ \bibinfo {pages} {074002} (\bibinfo {year} {2010})}\BibitemShut {NoStop}%
\bibitem [{\citenamefont {Ramos}\ \emph {et~al.}(2017)\citenamefont {Ramos}, \citenamefont {Moore},\ and\ \citenamefont {Raithel}}]{circ3}%
  \BibitemOpen
  \bibfield  {author} {\bibinfo {author} {\bibfnamefont {A.}~\bibnamefont {Ramos}}, \bibinfo {author} {\bibfnamefont {K.}~\bibnamefont {Moore}}, \ and\ \bibinfo {author} {\bibfnamefont {G.}~\bibnamefont {Raithel}},\ }\href {\doibase 10.1103/PhysRevA.96.032513} {\bibfield  {journal} {\bibinfo  {journal} {Phys. Rev. A}\ }\textbf {\bibinfo {volume} {96}},\ \bibinfo {pages} {032513} (\bibinfo {year} {2017})}\BibitemShut {NoStop}%
\bibitem [{\citenamefont {Safronova}\ \emph {et~al.}(1999)\citenamefont {Safronova}, \citenamefont {Johnson},\ and\ \citenamefont {Derevianko}}]{IonicTh1}%
  \BibitemOpen
  \bibfield  {author} {\bibinfo {author} {\bibfnamefont {M.~S.}\ \bibnamefont {Safronova}}, \bibinfo {author} {\bibfnamefont {W.~R.}\ \bibnamefont {Johnson}}, \ and\ \bibinfo {author} {\bibfnamefont {A.}~\bibnamefont {Derevianko}},\ }\href {\doibase 10.1103/PhysRevA.60.4476} {\bibfield  {journal} {\bibinfo  {journal} {Phys. Rev. A}\ }\textbf {\bibinfo {volume} {60}},\ \bibinfo {pages} {4476} (\bibinfo {year} {1999})}\BibitemShut {NoStop}%
\bibitem [{\citenamefont {Holmgren}\ \emph {et~al.}(2012)\citenamefont {Holmgren}, \citenamefont {Trubko}, \citenamefont {Hromada},\ and\ \citenamefont {Cronin}}]{TuneOut}%
  \BibitemOpen
  \bibfield  {author} {\bibinfo {author} {\bibfnamefont {W.~F.}\ \bibnamefont {Holmgren}}, \bibinfo {author} {\bibfnamefont {R.}~\bibnamefont {Trubko}}, \bibinfo {author} {\bibfnamefont {I.}~\bibnamefont {Hromada}}, \ and\ \bibinfo {author} {\bibfnamefont {A.~D.}\ \bibnamefont {Cronin}},\ }\href {\doibase 10.1103/PhysRevLett.109.243004} {\bibfield  {journal} {\bibinfo  {journal} {Phys. Rev. Lett.}\ }\textbf {\bibinfo {volume} {109}},\ \bibinfo {pages} {243004} (\bibinfo {year} {2012})}\BibitemShut {NoStop}%
\bibitem [{\citenamefont {Ratkata}\ \emph {et~al.}(2021)\citenamefont {Ratkata}, \citenamefont {Gregory}, \citenamefont {Innes}, \citenamefont {Matthies}, \citenamefont {McArd}, \citenamefont {Mortlock}, \citenamefont {Safronova}, \citenamefont {Bromley},\ and\ \citenamefont {Cornish}}]{TuneOutCs}%
  \BibitemOpen
  \bibfield  {author} {\bibinfo {author} {\bibfnamefont {A.}~\bibnamefont {Ratkata}}, \bibinfo {author} {\bibfnamefont {P.~D.}\ \bibnamefont {Gregory}}, \bibinfo {author} {\bibfnamefont {A.~D.}\ \bibnamefont {Innes}}, \bibinfo {author} {\bibfnamefont {A.~J.}\ \bibnamefont {Matthies}}, \bibinfo {author} {\bibfnamefont {L.~A.}\ \bibnamefont {McArd}}, \bibinfo {author} {\bibfnamefont {J.~M.}\ \bibnamefont {Mortlock}}, \bibinfo {author} {\bibfnamefont {M.~S.}\ \bibnamefont {Safronova}}, \bibinfo {author} {\bibfnamefont {S.~L.}\ \bibnamefont {Bromley}}, \ and\ \bibinfo {author} {\bibfnamefont {S.~L.}\ \bibnamefont {Cornish}},\ }\href {\doibase 10.1103/PhysRevA.104.052813} {\bibfield  {journal} {\bibinfo  {journal} {Phys. Rev. A}\ }\textbf {\bibinfo {volume} {104}},\ \bibinfo {pages} {052813} (\bibinfo {year} {2021})}\BibitemShut {NoStop}%
\bibitem [{\citenamefont {Mohapatra}\ \emph {et~al.}(2007)\citenamefont {Mohapatra}, \citenamefont {Jackson},\ and\ \citenamefont {Adams}}]{eit}%
  \BibitemOpen
  \bibfield  {author} {\bibinfo {author} {\bibfnamefont {A.~K.}\ \bibnamefont {Mohapatra}}, \bibinfo {author} {\bibfnamefont {T.~R.}\ \bibnamefont {Jackson}}, \ and\ \bibinfo {author} {\bibfnamefont {C.~S.}\ \bibnamefont {Adams}},\ }\href {\doibase 10.1103/PhysRevLett.98.113003} {\bibfield  {journal} {\bibinfo  {journal} {Phys. Rev. Lett.}\ }\textbf {\bibinfo {volume} {98}},\ \bibinfo {pages} {113003} (\bibinfo {year} {2007})}\BibitemShut {NoStop}%
\bibitem [{\citenamefont {Carr}\ \emph {et~al.}(2012)\citenamefont {Carr}, \citenamefont {Adams},\ and\ \citenamefont {Weatherill}}]{Carr:12}%
  \BibitemOpen
  \bibfield  {author} {\bibinfo {author} {\bibfnamefont {C.}~\bibnamefont {Carr}}, \bibinfo {author} {\bibfnamefont {C.~S.}\ \bibnamefont {Adams}}, \ and\ \bibinfo {author} {\bibfnamefont {K.~J.}\ \bibnamefont {Weatherill}},\ }\href {\doibase 10.1364/OL.37.000118} {\bibfield  {journal} {\bibinfo  {journal} {Opt. Lett.}\ }\textbf {\bibinfo {volume} {37}},\ \bibinfo {pages} {118} (\bibinfo {year} {2012})}\BibitemShut {NoStop}%
\bibitem [{\citenamefont {Chopinaud}\ and\ \citenamefont {Pritchard}(2021)}]{pritchard2021}%
  \BibitemOpen
  \bibfield  {author} {\bibinfo {author} {\bibfnamefont {A.}~\bibnamefont {Chopinaud}}\ and\ \bibinfo {author} {\bibfnamefont {J.}~\bibnamefont {Pritchard}},\ }\href {\doibase 10.1103/PhysRevApplied.16.024008} {\bibfield  {journal} {\bibinfo  {journal} {Phys. Rev. Appl.}\ }\textbf {\bibinfo {volume} {16}},\ \bibinfo {pages} {024008} (\bibinfo {year} {2021})}\BibitemShut {NoStop}%
\bibitem [{\citenamefont {Allinson}\ \emph {et~al.}(2024)\citenamefont {Allinson}, \citenamefont {Jamieson}, \citenamefont {Mackellar}, \citenamefont {Downes}, \citenamefont {Adams},\ and\ \citenamefont {Weatherill}}]{allinson2024simultaneous}%
  \BibitemOpen
  \bibfield  {author} {\bibinfo {author} {\bibfnamefont {G.}~\bibnamefont {Allinson}}, \bibinfo {author} {\bibfnamefont {M.~J.}\ \bibnamefont {Jamieson}}, \bibinfo {author} {\bibfnamefont {A.~R.}\ \bibnamefont {Mackellar}}, \bibinfo {author} {\bibfnamefont {L.}~\bibnamefont {Downes}}, \bibinfo {author} {\bibfnamefont {C.~S.}\ \bibnamefont {Adams}}, \ and\ \bibinfo {author} {\bibfnamefont {K.~J.}\ \bibnamefont {Weatherill}},\ }\href@noop {} {\bibfield  {journal} {\bibinfo  {journal} {Physical Review Research}\ }\textbf {\bibinfo {volume} {6}},\ \bibinfo {pages} {023317} (\bibinfo {year} {2024})}\BibitemShut {NoStop}%
\bibitem [{\citenamefont {Martin}(1980)}]{martin1980series}%
  \BibitemOpen
  \bibfield  {author} {\bibinfo {author} {\bibfnamefont {W.}~\bibnamefont {Martin}},\ }\href@noop {} {\bibfield  {journal} {\bibinfo  {journal} {JOSA}\ }\textbf {\bibinfo {volume} {70}},\ \bibinfo {pages} {784} (\bibinfo {year} {1980})}\BibitemShut {NoStop}%
\bibitem [{\citenamefont {Hughes}\ and\ \citenamefont {Hase}(2010)}]{Hughes:10}%
  \BibitemOpen
  \bibfield  {author} {\bibinfo {author} {\bibfnamefont {I.~G.}\ \bibnamefont {Hughes}}\ and\ \bibinfo {author} {\bibfnamefont {T.~P.~A.}\ \bibnamefont {Hase}},\ }\href@noop {} {\emph {\bibinfo {title} {Measurements and their Uncertainties}}}\ (\bibinfo  {publisher} {Oxford University Press},\ \bibinfo {year} {2010})\BibitemShut {NoStop}%
\bibitem [{http://doi.org/10.15128/r2df65v802x()}]{data_doi}%
  \BibitemOpen
  http://doi.org/10.15128/r2df65v802x,\ \href@noop {} {}\BibitemShut {NoStop}%
\bibitem [{\citenamefont {Berl}\ \emph {et~al.}(2020)\citenamefont {Berl}, \citenamefont {Sackett}, \citenamefont {Gallagher},\ and\ \citenamefont {Nunkaew}}]{berl}%
  \BibitemOpen
  \bibfield  {author} {\bibinfo {author} {\bibfnamefont {S.~J.}\ \bibnamefont {Berl}}, \bibinfo {author} {\bibfnamefont {C.~A.}\ \bibnamefont {Sackett}}, \bibinfo {author} {\bibfnamefont {T.~F.}\ \bibnamefont {Gallagher}}, \ and\ \bibinfo {author} {\bibfnamefont {J.}~\bibnamefont {Nunkaew}},\ }\href {\doibase 10.1103/PhysRevA.102.062818} {\bibfield  {journal} {\bibinfo  {journal} {Phys. Rev. A}\ }\textbf {\bibinfo {volume} {102}},\ \bibinfo {pages} {062818} (\bibinfo {year} {2020})}\BibitemShut {NoStop}%
\bibitem [{\citenamefont {Safinya}\ \emph {et~al.}(1980)\citenamefont {Safinya}, \citenamefont {Gallagher},\ and\ \citenamefont {Sandner}}]{safinya1980resonance}%
  \BibitemOpen
  \bibfield  {author} {\bibinfo {author} {\bibfnamefont {K.}~\bibnamefont {Safinya}}, \bibinfo {author} {\bibfnamefont {T.}~\bibnamefont {Gallagher}}, \ and\ \bibinfo {author} {\bibfnamefont {W.}~\bibnamefont {Sandner}},\ }\href@noop {} {\bibfield  {journal} {\bibinfo  {journal} {Physical Review A}\ }\textbf {\bibinfo {volume} {22}},\ \bibinfo {pages} {2672} (\bibinfo {year} {1980})}\BibitemShut {NoStop}%
\bibitem [{\citenamefont {Komara}\ \emph {et~al.}(2003)\citenamefont {Komara}, \citenamefont {Gearba}, \citenamefont {Lundeen},\ and\ \citenamefont {Fehrenbach}}]{Silicon}%
  \BibitemOpen
  \bibfield  {author} {\bibinfo {author} {\bibfnamefont {R.~A.}\ \bibnamefont {Komara}}, \bibinfo {author} {\bibfnamefont {M.~A.}\ \bibnamefont {Gearba}}, \bibinfo {author} {\bibfnamefont {S.~R.}\ \bibnamefont {Lundeen}}, \ and\ \bibinfo {author} {\bibfnamefont {C.~W.}\ \bibnamefont {Fehrenbach}},\ }\href {\doibase 10.1103/PhysRevA.67.062502} {\bibfield  {journal} {\bibinfo  {journal} {Phys. Rev. A}\ }\textbf {\bibinfo {volume} {67}},\ \bibinfo {pages} {062502} (\bibinfo {year} {2003})}\BibitemShut {NoStop}%
\bibitem [{\citenamefont {Sansonetti}\ \emph {et~al.}(1981)\citenamefont {Sansonetti}, \citenamefont {Andrew},\ and\ \citenamefont {Verges}}]{Sansonetti:81}%
  \BibitemOpen
  \bibfield  {author} {\bibinfo {author} {\bibfnamefont {C.~J.}\ \bibnamefont {Sansonetti}}, \bibinfo {author} {\bibfnamefont {K.~L.}\ \bibnamefont {Andrew}}, \ and\ \bibinfo {author} {\bibfnamefont {J.}~\bibnamefont {Verges}},\ }\href {\doibase 10.1364/JOSA.71.000423} {\bibfield  {journal} {\bibinfo  {journal} {J. Opt. Soc. Am.}\ }\textbf {\bibinfo {volume} {71}},\ \bibinfo {pages} {423} (\bibinfo {year} {1981})}\BibitemShut {NoStop}%
\bibitem [{\citenamefont {Patil}(1994)}]{Patil_1994}%
  \BibitemOpen
  \bibfield  {author} {\bibinfo {author} {\bibfnamefont {S.~H.}\ \bibnamefont {Patil}},\ }\href {\doibase 10.1088/0953-4075/27/3/024} {\bibfield  {journal} {\bibinfo  {journal} {Journal of Physics B: Atomic, Molecular and Optical Physics}\ }\textbf {\bibinfo {volume} {27}},\ \bibinfo {pages} {601} (\bibinfo {year} {1994})}\BibitemShut {NoStop}%
\bibitem [{\citenamefont {Drachman}(1982{\natexlab{b}})}]{drachman1982rydberg}%
  \BibitemOpen
  \bibfield  {author} {\bibinfo {author} {\bibfnamefont {R.~J.}\ \bibnamefont {Drachman}},\ }\href@noop {} {\bibfield  {journal} {\bibinfo  {journal} {Physical Review A}\ }\textbf {\bibinfo {volume} {26}},\ \bibinfo {pages} {1228} (\bibinfo {year} {1982}{\natexlab{b}})}\BibitemShut {NoStop}%
\bibitem [{\citenamefont {Gallagher}(1994)}]{Gallagher:94}%
  \BibitemOpen
  \bibfield  {author} {\bibinfo {author} {\bibfnamefont {T.~F.}\ \bibnamefont {Gallagher}},\ }\href {\doibase 10.1017/CBO9780511524530} {\emph {\bibinfo {title} {Rydberg Atoms}}},\ Cambridge Monographs on Atomic, Molecular and Chemical Physics\ (\bibinfo  {publisher} {Cambridge University Press},\ \bibinfo {year} {1994})\BibitemShut {NoStop}%
\bibitem [{\citenamefont {Bockasten}(1974)}]{Bock}%
  \BibitemOpen
  \bibfield  {author} {\bibinfo {author} {\bibfnamefont {K.}~\bibnamefont {Bockasten}},\ }\href {\doibase 10.1103/PhysRevA.9.1087} {\bibfield  {journal} {\bibinfo  {journal} {Phys. Rev. A}\ }\textbf {\bibinfo {volume} {9}},\ \bibinfo {pages} {1087} (\bibinfo {year} {1974})}\BibitemShut {NoStop}%
\bibitem [{\citenamefont {Curtis}\ and\ \citenamefont {Ramanujam}(1981)}]{Curtis:81}%
  \BibitemOpen
  \bibfield  {author} {\bibinfo {author} {\bibfnamefont {L.~J.}\ \bibnamefont {Curtis}}\ and\ \bibinfo {author} {\bibfnamefont {P.~S.}\ \bibnamefont {Ramanujam}},\ }\href {\doibase 10.1364/JOSA.71.001315} {\bibfield  {journal} {\bibinfo  {journal} {J. Opt. Soc. Am.}\ }\textbf {\bibinfo {volume} {71}},\ \bibinfo {pages} {1315} (\bibinfo {year} {1981})}\BibitemShut {NoStop}%
\bibitem [{\citenamefont {Eriksson}\ and\ \citenamefont {Wenåker}(1970)}]{Eriksson_1970}%
  \BibitemOpen
  \bibfield  {author} {\bibinfo {author} {\bibfnamefont {K.~B.~S.}\ \bibnamefont {Eriksson}}\ and\ \bibinfo {author} {\bibfnamefont {I.}~\bibnamefont {Wenåker}},\ }\href {\doibase 10.1088/0031-8949/1/1/003} {\bibfield  {journal} {\bibinfo  {journal} {Physica Scripta}\ }\textbf {\bibinfo {volume} {1}},\ \bibinfo {pages} {21} (\bibinfo {year} {1970})}\BibitemShut {NoStop}%
\bibitem [{\citenamefont {Fredriksson}\ \emph {et~al.}(1980)\citenamefont {Fredriksson}, \citenamefont {Lundberg},\ and\ \citenamefont {Svanberg}}]{Fredrissin1980}%
  \BibitemOpen
  \bibfield  {author} {\bibinfo {author} {\bibfnamefont {K.}~\bibnamefont {Fredriksson}}, \bibinfo {author} {\bibfnamefont {H.}~\bibnamefont {Lundberg}}, \ and\ \bibinfo {author} {\bibfnamefont {S.}~\bibnamefont {Svanberg}},\ }\href {\doibase 10.1103/PhysRevA.21.241} {\bibfield  {journal} {\bibinfo  {journal} {Phys. Rev. A}\ }\textbf {\bibinfo {volume} {21}},\ \bibinfo {pages} {241} (\bibinfo {year} {1980})}\BibitemShut {NoStop}%
\bibitem [{\citenamefont {Eissa}\ and\ \citenamefont {Öpik}(1967)}]{Eissa_1967}%
  \BibitemOpen
  \bibfield  {author} {\bibinfo {author} {\bibfnamefont {H.}~\bibnamefont {Eissa}}\ and\ \bibinfo {author} {\bibfnamefont {U.}~\bibnamefont {Öpik}},\ }\href {\doibase 10.1088/0370-1328/92/3/307} {\bibfield  {journal} {\bibinfo  {journal} {Proceedings of the Physical Society}\ }\textbf {\bibinfo {volume} {92}},\ \bibinfo {pages} {556} (\bibinfo {year} {1967})}\BibitemShut {NoStop}%
\bibitem [{\citenamefont {Drake}\ and\ \citenamefont {Swainson}(1991)}]{DrakeSwan}%
  \BibitemOpen
  \bibfield  {author} {\bibinfo {author} {\bibfnamefont {G.~W.~F.}\ \bibnamefont {Drake}}\ and\ \bibinfo {author} {\bibfnamefont {R.~A.}\ \bibnamefont {Swainson}},\ }\href {\doibase 10.1103/PhysRevA.44.5448} {\bibfield  {journal} {\bibinfo  {journal} {Phys. Rev. A}\ }\textbf {\bibinfo {volume} {44}},\ \bibinfo {pages} {5448} (\bibinfo {year} {1991})}\BibitemShut {NoStop}%
\bibitem [{\citenamefont {Kleinman}\ \emph {et~al.}(1968)\citenamefont {Kleinman}, \citenamefont {Hahn},\ and\ \citenamefont {Spruch}}]{DomNA}%
  \BibitemOpen
  \bibfield  {author} {\bibinfo {author} {\bibfnamefont {C.~J.}\ \bibnamefont {Kleinman}}, \bibinfo {author} {\bibfnamefont {Y.}~\bibnamefont {Hahn}}, \ and\ \bibinfo {author} {\bibfnamefont {L.}~\bibnamefont {Spruch}},\ }\href {\doibase 10.1103/PhysRev.165.53} {\bibfield  {journal} {\bibinfo  {journal} {Phys. Rev.}\ }\textbf {\bibinfo {volume} {165}},\ \bibinfo {pages} {53} (\bibinfo {year} {1968})}\BibitemShut {NoStop}%
\bibitem [{\citenamefont {Johnson}\ \emph {et~al.}(1983)\citenamefont {Johnson}, \citenamefont {Kolb},\ and\ \citenamefont {Huang}}]{JOHNSON1983333}%
  \BibitemOpen
  \bibfield  {author} {\bibinfo {author} {\bibfnamefont {W.}~\bibnamefont {Johnson}}, \bibinfo {author} {\bibfnamefont {D.}~\bibnamefont {Kolb}}, \ and\ \bibinfo {author} {\bibfnamefont {K.-N.}\ \bibnamefont {Huang}},\ }\href {\doibase https://doi.org/10.1016/0092-640X(83)90020-7} {\bibfield  {journal} {\bibinfo  {journal} {Atomic Data and Nuclear Data Tables}\ }\textbf {\bibinfo {volume} {28}},\ \bibinfo {pages} {333} (\bibinfo {year} {1983})}\BibitemShut {NoStop}%
\bibitem [{\citenamefont {Lim}\ \emph {et~al.}(2002)\citenamefont {Lim}, \citenamefont {Laerdahl},\ and\ \citenamefont {Schwerdtfeger}}]{lim2002fully}%
  \BibitemOpen
  \bibfield  {author} {\bibinfo {author} {\bibfnamefont {I.~S.}\ \bibnamefont {Lim}}, \bibinfo {author} {\bibfnamefont {J.~K.}\ \bibnamefont {Laerdahl}}, \ and\ \bibinfo {author} {\bibfnamefont {P.}~\bibnamefont {Schwerdtfeger}},\ }\href@noop {} {\bibfield  {journal} {\bibinfo  {journal} {The Journal of chemical physics}\ }\textbf {\bibinfo {volume} {116}},\ \bibinfo {pages} {172} (\bibinfo {year} {2002})}\BibitemShut {NoStop}%
\bibitem [{\citenamefont {Safronova}\ \emph {et~al.}(2016)\citenamefont {Safronova}, \citenamefont {Safronova},\ and\ \citenamefont {Clark}}]{newCs}%
  \BibitemOpen
  \bibfield  {author} {\bibinfo {author} {\bibfnamefont {M.~S.}\ \bibnamefont {Safronova}}, \bibinfo {author} {\bibfnamefont {U.~I.}\ \bibnamefont {Safronova}}, \ and\ \bibinfo {author} {\bibfnamefont {C.~W.}\ \bibnamefont {Clark}},\ }\href {\doibase 10.1103/PhysRevA.94.012505} {\bibfield  {journal} {\bibinfo  {journal} {Phys. Rev. A}\ }\textbf {\bibinfo {volume} {94}},\ \bibinfo {pages} {012505} (\bibinfo {year} {2016})}\BibitemShut {NoStop}%
\bibitem [{\citenamefont {Sternheimer}(1970)}]{sternheimer1970quadrupole}%
  \BibitemOpen
  \bibfield  {author} {\bibinfo {author} {\bibfnamefont {R.}~\bibnamefont {Sternheimer}},\ }\href@noop {} {\bibfield  {journal} {\bibinfo  {journal} {Physical Review A}\ }\textbf {\bibinfo {volume} {1}},\ \bibinfo {pages} {321} (\bibinfo {year} {1970})}\BibitemShut {NoStop}%
\bibitem [{\citenamefont {Mahan}(1980)}]{mahan1980modified}%
  \BibitemOpen
  \bibfield  {author} {\bibinfo {author} {\bibfnamefont {G.}~\bibnamefont {Mahan}},\ }\href@noop {} {\bibfield  {journal} {\bibinfo  {journal} {Physical Review A}\ }\textbf {\bibinfo {volume} {22}},\ \bibinfo {pages} {1780} (\bibinfo {year} {1980})}\BibitemShut {NoStop}%
\bibitem [{\citenamefont {Weber}\ \emph {et~al.}(2017)\citenamefont {Weber}, \citenamefont {Tresp}, \citenamefont {Menke}, \citenamefont {Urvoy}, \citenamefont {Firstenberg}, \citenamefont {Büchler},\ and\ \citenamefont {Hofferberth}}]{Weberint}%
  \BibitemOpen
  \bibfield  {author} {\bibinfo {author} {\bibfnamefont {S.}~\bibnamefont {Weber}}, \bibinfo {author} {\bibfnamefont {C.}~\bibnamefont {Tresp}}, \bibinfo {author} {\bibfnamefont {H.}~\bibnamefont {Menke}}, \bibinfo {author} {\bibfnamefont {A.}~\bibnamefont {Urvoy}}, \bibinfo {author} {\bibfnamefont {O.}~\bibnamefont {Firstenberg}}, \bibinfo {author} {\bibfnamefont {H.~P.}\ \bibnamefont {Büchler}}, \ and\ \bibinfo {author} {\bibfnamefont {S.}~\bibnamefont {Hofferberth}},\ }\href {\doibase 10.1088/1361-6455/aa743a} {\bibfield  {journal} {\bibinfo  {journal} {Journal of Physics B: Atomic, Molecular and Optical Physics}\ }\textbf {\bibinfo {volume} {50}},\ \bibinfo {pages} {133001} (\bibinfo {year} {2017})}\BibitemShut {NoStop}%
\bibitem [{\citenamefont {Greene}\ \emph {et~al.}(2000)\citenamefont {Greene}, \citenamefont {Dickinson},\ and\ \citenamefont {Sadeghpour}}]{ulr_mol}%
  \BibitemOpen
  \bibfield  {author} {\bibinfo {author} {\bibfnamefont {C.~H.}\ \bibnamefont {Greene}}, \bibinfo {author} {\bibfnamefont {A.~S.}\ \bibnamefont {Dickinson}}, \ and\ \bibinfo {author} {\bibfnamefont {H.~R.}\ \bibnamefont {Sadeghpour}},\ }\href {\doibase 10.1103/PhysRevLett.85.2458} {\bibfield  {journal} {\bibinfo  {journal} {Phys. Rev. Lett.}\ }\textbf {\bibinfo {volume} {85}},\ \bibinfo {pages} {2458} (\bibinfo {year} {2000})}\BibitemShut {NoStop}%
\bibitem [{\citenamefont {Eiles}(2019)}]{Eiles_2019}%
  \BibitemOpen
  \bibfield  {author} {\bibinfo {author} {\bibfnamefont {M.~T.}\ \bibnamefont {Eiles}},\ }\href {\doibase 10.1088/1361-6455/ab19ca} {\bibfield  {journal} {\bibinfo  {journal} {Journal of Physics B: Atomic, Molecular and Optical Physics}\ }\textbf {\bibinfo {volume} {52}},\ \bibinfo {pages} {113001} (\bibinfo {year} {2019})}\BibitemShut {NoStop}%
\bibitem [{\citenamefont {Shen}\ \emph {et~al.}(2024)\citenamefont {Shen}, \citenamefont {Booth}, \citenamefont {Liu}, \citenamefont {Beattie}, \citenamefont {Marceau}, \citenamefont {Shaffer}, \citenamefont {Pawlak},\ and\ \citenamefont {Sadeghpour}}]{Schaffer}%
  \BibitemOpen
  \bibfield  {author} {\bibinfo {author} {\bibfnamefont {P.}~\bibnamefont {Shen}}, \bibinfo {author} {\bibfnamefont {D.}~\bibnamefont {Booth}}, \bibinfo {author} {\bibfnamefont {C.}~\bibnamefont {Liu}}, \bibinfo {author} {\bibfnamefont {S.}~\bibnamefont {Beattie}}, \bibinfo {author} {\bibfnamefont {C.}~\bibnamefont {Marceau}}, \bibinfo {author} {\bibfnamefont {J.~P.}\ \bibnamefont {Shaffer}}, \bibinfo {author} {\bibfnamefont {M.}~\bibnamefont {Pawlak}}, \ and\ \bibinfo {author} {\bibfnamefont {H.~R.}\ \bibnamefont {Sadeghpour}},\ }\href {\doibase 10.1103/PhysRevLett.133.233005} {\bibfield  {journal} {\bibinfo  {journal} {Phys. Rev. Lett.}\ }\textbf {\bibinfo {volume} {133}},\ \bibinfo {pages} {233005} (\bibinfo {year} {2024})},\ \Eprint {http://arxiv.org/abs/2410.19129} {arXiv:2410.19129 [physics.atom-ph]} \BibitemShut {NoStop}%
\bibitem [{\citenamefont {Gray}\ \emph {et~al.}(1988{\natexlab{b}})\citenamefont {Gray}, \citenamefont {Sun},\ and\ \citenamefont {MacAdam}}]{Sodium2}%
  \BibitemOpen
  \bibfield  {author} {\bibinfo {author} {\bibfnamefont {L.~G.}\ \bibnamefont {Gray}}, \bibinfo {author} {\bibfnamefont {X.}~\bibnamefont {Sun}}, \ and\ \bibinfo {author} {\bibfnamefont {K.~B.}\ \bibnamefont {MacAdam}},\ }\href {\doibase 10.1103/PhysRevA.38.4985} {\bibfield  {journal} {\bibinfo  {journal} {Phys. Rev. A}\ }\textbf {\bibinfo {volume} {38}},\ \bibinfo {pages} {4985} (\bibinfo {year} {1988}{\natexlab{b}})}\BibitemShut {NoStop}%
\bibitem [{\citenamefont {Wade}\ \emph {et~al.}(2017)\citenamefont {Wade}, \citenamefont {\ifmmode \check{S}\else \v{S}\fi{}ibali\ifmmode~\acute{c}\else \'{c}\fi{}}, \citenamefont {de~Melo}, \citenamefont {Kondo}, \citenamefont {Weatherill},\ and\ \citenamefont {Adams}}]{Wade:17}%
  \BibitemOpen
  \bibfield  {author} {\bibinfo {author} {\bibfnamefont {C.~G.}\ \bibnamefont {Wade}}, \bibinfo {author} {\bibfnamefont {N.}~\bibnamefont {\ifmmode \check{S}\else \v{S}\fi{}ibali\ifmmode~\acute{c}\else \'{c}\fi{}}}, \bibinfo {author} {\bibfnamefont {N.~R.}\ \bibnamefont {de~Melo}}, \bibinfo {author} {\bibfnamefont {J.~M.}\ \bibnamefont {Kondo}}, \bibinfo {author} {\bibfnamefont {K.~J.}\ \bibnamefont {Weatherill}}, \ and\ \bibinfo {author} {\bibfnamefont {C.}~\bibnamefont {Adams}},\ }\href@noop {} {\bibfield  {journal} {\bibinfo  {journal} {Nature Photonics}\ }\textbf {\bibinfo {volume} {72}},\ \bibinfo {pages} {40} (\bibinfo {year} {2017})}\BibitemShut {NoStop}%
\bibitem [{\citenamefont {Downes}\ \emph {et~al.}(2023)\citenamefont {Downes}, \citenamefont {Torralbo-Campo},\ and\ \citenamefont {Weatherill}}]{Downes_2023}%
  \BibitemOpen
  \bibfield  {author} {\bibinfo {author} {\bibfnamefont {L.~A.}\ \bibnamefont {Downes}}, \bibinfo {author} {\bibfnamefont {L.}~\bibnamefont {Torralbo-Campo}}, \ and\ \bibinfo {author} {\bibfnamefont {K.~J.}\ \bibnamefont {Weatherill}},\ }\href {\doibase 10.1088/1367-2630/acb80c} {\bibfield  {journal} {\bibinfo  {journal} {New Journal of Physics}\ }\textbf {\bibinfo {volume} {25}},\ \bibinfo {pages} {035002} (\bibinfo {year} {2023})}\BibitemShut {NoStop}%
\end{thebibliography}%

\end{document}